\documentclass[aps,prd,superscriptaddress,nofootinbib,amsmath,amsfonts,preprintnumbers,groupedaddress,showpacs,10pt,english]{revtex4-1}
\usepackage{amsmath}
\usepackage{amssymb}
\usepackage{babel}
\usepackage{wrapfig}
\usepackage{cancel}
\newcommand{\nn}{\nonumber \\}
\newcommand{\e}{\mathrm{e}}
\usepackage{relsize,exscale}
\makeatletter

\usepackage{array,multirow,graphicx}
\usepackage{dcolumn}
\usepackage{newlfont}
\usepackage{bm}
\usepackage[colorlinks,citecolor=blue,urlcolor=blue,linkcolor=blue]{hyperref}
\usepackage[figtopcap]{subfigure}
\usepackage{color}

\allowdisplaybreaks[4]

\begin{document}

\tolerance=5000

\date{\today}

\title{Specific neutral and charged black holes in $f(R)$ gravitational theory}

\author{G.~G.~L.~Nashed}
\email{nashed@bue.edu.eg}
\affiliation {Centre for Theoretical Physics, The British University, P.O. Box
43, El Sherouk City, Cairo 11837, Egypt}
\author{Shin'ichi~Nojiri}
\email{nojiri@gravity.phys.nagoya-u.ac.jp}
\affiliation{Department of Physics, Nagoya University, Nagoya 464-8602 \\
\& \\
Japan Kobayashi-Maskawa Institute for the Origin of Particles and the Universe,
Nagoya University, Nagoya 464-8602, Japan }

\begin{abstract}
With the successes of $f(R)$ theory as a neutral modification of
Einstein's general relativity (GR), we continue our study
in this field and attempt to find general 
{ neutral}
and charged black hole (BH) solutions.
In the previous papers \cite{Nashed:2020mnp,Nashed:2020tbp},
we applied the field equation of the $f(R)$ gravity
to a spherically symmetric space-time
$ds^2=-U(r)dt^2+\frac{dr^2}{V(r)}+r^2 \left( d\theta^2+\sin^2\theta d\phi^2 \right)$
with unequal metric potentials $U(r)$ and $V(r)$ and with/without electric charge.
{Then we have obtained equations which include all the possible static solutions
with spherical symmetry.}
To ensure the closed form of system of the resulting differential equations in order to obtain specific solutions,
we assumed the derivative of the $f(R)$ with respect to the scalar curvature $R$ to have a form
$F_1(r) {=\frac{df(R(r))}{dR(r)} } = 1-\frac{F_0-\left(n-3\right)}{r^n}$ with a constant $F_0$
and show that we can generate asymptotically GR BH solutions for $n>2$ but we show that the $n=2$ case is not allowed.
This form of $F_1(r)$ could be the most acceptable physical form that we can generate from it physical metric potentials
that can have a well-known asymptotic form and we obtain the metric of the Einstein general relativity in the limit of $F_0\to n-3$.
We show that the form of the electric charge depends on $n$ and that $n\neq 2$.
Our study shows that the power $n$ is sensitive and why we should exclude the case $n=2$ for the choice of $F_1(r)$ presented in this study.
We also study the physics of these black hole solutions by calculating their thermodynamical quantities, like entropy, the Hawking temperature
and Gibb's free energy, and derive the stability conditions by using geodesic deviations.
{ In the standard Reissner-Nordstr\"om space-time which is the charged black hole solution in GR,
there appear two black hole horizons, that is, inner horizon and outer horizon.
When the radii of the two horizons coincide with each other, which is called the extremal limit,
the absolute value of the charge equals to the mass and the Hawking temperature vanishes.
In our model, however, the absolute value of the charge is not equal to the mass in the limit
althought the Hawking temperature vanishes. }

\end{abstract}

\maketitle
\section{Introduction}\label{Sec1}

The $f(R)$ gravity is a modified gravitational theory where the action is given in a generic { function} 
of scalar curvature, $R$ and the $f(R)$ gravity may describe the early and late cosmological evolution.
The $f(R)$ theory can describe dark energy and dark matter consistently without imposing any new material
that has not yet been revealed by experiments \cite{Nojiri:2010wj,Nojiri:2006ri,Copeland:2006wr,Aragon:2020xtm,Clifton:2011jh,Capozziello:2007ec}.
Amending the Lagrangian of general relativity (GR) does not only influence the dynamical system of the universe but
it can also change the dynamical system at the galactic or solar system scales.
Thus, amended theories of gravity with higher-order curvature corrections provide a deeper understanding of gravity.


The Einstein-Hilbert action, which reproduces the field equation of GR, is linear to the Ricci scalar $R$.
By changing the action to include the non-linear terms, the Ricci curvature, and/or Riemann curvatures,
many viable modified gravitational theories are presented by the scientific society to describe the cosmic evolution in early times.
Most of those theories use a gravitational Lagrangian which contains some of the four possible second-order curvature invariants.
Moreover, many models that use higher-order invariants as a function of $R$ are introduced in the gravitational action
and different $f(R)$ gravitational models are obtained \cite{Mohsenzadeh:2012ka, Cognola:2007zu,Awad:2017ign,Pogosian:2007sw,Zhang:2005vt,
Li:2007xn,Song:2007da,Nojiri:2007cq,2017JHEP...07..136A,Nojiri:2007as,Capozziello:2018ddp,Nashed:2006yw,Vainio:2016qas}.
Aside from the ability of these theories to eliminate the contributions of curvature invariants other than the Ricci scalar $R$,
they could also prevent the Ostrogradski issue \cite{Ostrogradsky:1850fid}, which is
a problem that characterizes any higher derivative theories \cite{Woodard:2006nt}.

The earliest modification of GR could be the one presented in \cite{Buchdahl:1983zz}.
A natural extension of GR is to include expressions such as $R^n$ with a constant $n$ similar to the Starobinsky model
$f(R) = R + \epsilon R^2$, where $\epsilon$ is also a constant \cite{Starobinsky:1980te}.
When $n<0$, the contribution of $R^n$ could investigate the late epoch and describe self-accelerating vacuum solutions
\cite{Carroll:2003st,Carroll:2003wy,Capozziello:2002rd,Capozziello:2003gx}.
Nevertheless, such solutions suffer from instabilities \cite{Soussa:2003re,Faraoni:2005vk}
and have strong limitations from the solar system test \cite{Chiba:2003ir}.
To avoid the previously mentioned issues, scientists have considered the $f(R)$ gravitational theory,
which can accommodate a wide range of phenomena.
Many applications carried use the  framework of the $f(R)$ gravitational theory such as
gravitational wave detection \cite{Corda:2008si,Corda:2010zza}, early-time
inflation \cite{Bamba:2008ja}, cosmological phases \cite{Nojiri:2006gh,Nashed:2018oaf,Nojiri:2006be,Nojiri:2008fk},
the singularity problem \cite{Kobayashi:2008wc}, stability of solutions \cite{Faraoni:2005ie,Capozziello:2006dj,Amendola:2006we},
and other different issues \cite{Akbar:2006mq}.

Many black hole (BH) solutions in the $f(R)$ theory coincide with the BH solutions of GR or differ from them.
Among these solutions, the authors in \cite{Multamaki:2006zb} derived static spherically symmetric solutions
and showed that the de Sitter(dS)-Schwarzschild metric is a solution to the field equations of the $f(R)$ theory.
Moreover, spherically symmetric solutions are derived in \cite{Multamaki:2006ym} by involving a perfect fluid.
The BH solution with/without electric charge is presented in \cite{delaCruzDombriz:2012xy}.
Many analytic spherically symmetric solutions are derived in \cite{Hendi:2011eg,Nashed:2019uyi,Sebastiani:2010kv,
Nashed:2004pn,Hendi:2012nj,Nashed:2019ykm,Nashed:2018piz,Nashed:2018efg,
Asgari:2011ix,Nashed:2009hn,Ghosh:2014hea,Hendi:2014mba}.
Analytic charged $D$-dimensional BH solutions are derived and discussed in \cite{Tang:2019qiy}.
Moreover, analytic spherically symmetric BH solutions with/without electric charge are derived
in \cite{Nashed:2020mnp,Nashed:2020tbp}.
Those BH solutions were different from the BH solutions of GR and coincide with them under certain special conditions.
The study in this paper aims to generalize this special condition and derive analytic BH solutions
with/without electric charge in the $f(R)$ gravitational theory.

This paper is organized as follows:
In Section~\ref{S2}, we give the building block of the $f(R)$ gravitational theory
and obtain its field equations including the contribution from the Maxwell field.
In Section~\ref{S3}, we apply the charged field equations of the $f(R)$ gravity
to a space-time having spherically symmetric and unequal metric potentials.
We present the non-linear differential equations which are composed
of five non-linear differential equations having four unknown functions, one is
the electric { potential,} 
one is the derivative $f'(R)$ of the $f(R)$,
and the other two are related to the metric potentials.
We study special cases that give a consistent BH solution which was derived in the previous literature.
Then we study the general case and divide it into two classes:
The first class is the one without charge and derive an original new BH solution
assuming the first derivative of $f(R)$ concerning $R$
to has the form $f'(R)=1-\frac{F_0-\left(n-3\right)}{r^n}$ where $n$ can be any value.
The second class is the one with the electric charge, and we derive a new charged BH solution assuming
that the derivative $f'(R)$ is not changed from that in the case without electric charge.
In Section~\ref{S4}, we study the physical properties of these new BH solutions with/without charge by giving
the form of the metric potentials in asymptotic form and show that they are different from GR metric potentials
either the Schwarzschild or the Reissner-Nordstr\"om {space-time}
due to the contribution of the non-linear curvature scalar terms.
We also study the 
{
scalar invariants made of curvatures} of the BH solution and show that its { singularity} 
is softer than that of GR for the case with an electric charge.
In Section~\ref{S6336b}, we present the stability constraints
of those BH solutions by using geodesic deviation and investigate the regions of stability graphically.
In Section~\ref{S5}, we evaluate the basic thermodynamical expressions, that is,
the Hawking-temperature, entropy, quasi-local energy,
heat capacity, and Gibb's free energy, related to our new BH solutions
and show that the solutions are physically acceptable.
{ Our charged BH  corresponds to the Reissner-Nordstr\"om space-time, which is the charged BH solution in GR.  In the solution, there appear two black hole horizons, which are called the inner horizon and the outer horizon.
The extremal limit is the limit when the radii of the two horizons coincide with each other. In the limit, the absolute value of the charge coincides with the mass, and the Hawking temperature vanishes.
In our model, however, the absolute value of the charge does not equal the mass in the limit
although the Hawking temperature vanishes. }
In subsection~\ref{fir}, we explain that the new BH solutions fulfill the first law of thermodynamics.
In the final section, we discuss our derived results.

\section{$f(R)$ amended theory}\label{S2}

If $f(R)\neq R$ then, we have a modified gravitational theory that is unlike GR.
The four dimensional action of the $f(R)$ gravitational theory takes the following form \cite{Carroll:2003wy,
1970MNRAS.150....1B,Nojiri:2003ft,Capozziello:2003gx,Capozziello:2011et,Nashed:2001im,Nojiri:2010wj,Nojiri:2017ncd,Capozziello:2002rd},
\begin{eqnarray}
\label{a2}
{\mathop{\mathcal{S}}}:=\frac{1}{2\kappa} \int d^4x \sqrt{-g} f(R)+\int d^{4}x~\sqrt{-g}~{\mathcal L}_\mathrm{em}\,,
\end{eqnarray}
with $g$ being the determinant of the metric and $\kappa$ being the Newtonian gravitational constant.
The Maxwell electromagnetic field Lagrangian ${\mathcal L}_\mathrm{em}$ is given by
${\mathcal{L}_\mathrm{em}=\frac{1}{4} F^2\equiv F_{\mu \nu}F^{\mu \nu}}$
where $F_{\mu\nu} = \partial_\mu\eta_\nu - \partial_\nu \eta_\mu$ and $\eta =\eta _\mu dx^\mu$
is the electromagnetic Maxwell gauge potential 1-form \cite{Capozziello:2012zj}.

Using the variations principle of the action (\ref{a2}), we obtain the field equations of the $f(R)$ gravity
in the following form \cite{2005JCAP...02..010C},
\begin{eqnarray}
\label{f1}
R_{\mu \nu} f_R-\frac{1}{2}g_{\mu \nu}f(R)
+ \left[ g_{\mu \nu}\Box -\nabla_\mu \nabla_\nu \right] f_R
=-\frac{1}{2}\kappa T^\mathrm{em}_{\mu\nu}\,,
\end{eqnarray}
where $\Box$ is the d'Alembertian operator, and $f_R=\frac{df}{dR}$ and
$T^\mathrm{em}_{\mu\nu}$ is the
energy-momentum tensor of the Maxwell field defined as
\begin{equation}
\label{en11}
T^{\mathrm{em}}_{\mu\nu}=F_{\mu \alpha}F_{\nu}^{\ \alpha}
 -\frac{1}{4} g_{\mu \nu} F^2\, .
\end{equation}
Furthermore, by the variation of equation (\ref{a2}) with respect to the gauge potential,
$\eta _{\mu}$, we obtain
\begin{equation}
\label{q8b}
{\partial_\nu\left( \sqrt{-g} F^{\mu \nu} \right)=0}\, .
\end{equation}
Taking the trace of equations~(\ref{f1}), we find
\begin{eqnarray}
\label{f3}
3\Box {f_R}+R{f_R}-2f(R)=0 \,.
\end{eqnarray}
Using Eq.~(\ref{f3}), we obtain the form of the $f(R)$ as follows,
\begin{eqnarray}
\label{f3s}
f(R)=\frac{1}{2}\left[ 3\Box f_{\mathrm {R}}+R{f_R} \right]\,.
\end{eqnarray}
 From Eqs.~(\ref{f3s}) and (\ref{f1}), we obtain \cite{Kalita:2019xjq}
\begin{eqnarray} \label{f3ss}
R_{\mu \nu} f_R-\frac{1}{4}g_{\mu \nu}Rf_R+\frac{1}{4}g_{\mu \nu}\Box f_R
-\nabla_\mu \nabla_\nu f_R
+\frac{1}{2}\kappa T^\mathrm{em}_{\mu\nu}=0 \,.
\end{eqnarray}
Therefore, a significant step is to test Eqs.~(\ref{q8b}) and (\ref{f3ss}) to a spherically symmetric space-time
whose line element has two different unknown functions.

\section{spherically symmetric BH solutions }\label{S3}

The spherically symmetric line-element is assumed to be given by
\begin{align}
\label{met12}
& & ds^2=-U(r)dt^2+\frac{dr^2}{V(r)}+r^2 d\Omega^2 \,,
\quad d\Omega^2= r^2 \left( d\theta^2+\sin^2d\phi^2 \right) \, ,
\end{align}
where $U(r)$ and $V(r)$ are two unknowns of $r$.
For the space-time (\ref{met12}), the Ricci scalar is evaluated as,
\begin{eqnarray}
\label{Ricci}
R(r)=\frac{r^2V{U'}^2-r^2UU'V'-2r^2UVU''-4rU \left[ VU'-UV' \right]+4U^2(1-V)}{2r^2U^2}\,,
\end{eqnarray}
where $U\equiv U(r)$, $V\equiv V(r)$, $U'=\frac{dU}{dr}$, $U''=\frac{d^2U}{dr^2}$, and $V'=\frac{dV}{dr}$.
Plugging Eqs.~(\ref{f3}) and (\ref{f3ss}) with Eq.~(\ref{met12}) and by using Eq.~(\ref{Ricci}), we obtain
the $(t,t)$, $(r,r)$, and $(\theta,\theta)$ (or $(\phi,\phi)$) components of the $f(R)$ equation
are given by
\begin{align}
\label{fftt}
0=&\frac{1}{8r^2W^2}\left\{r^2 \left[ V F_1{W'}^2-3W F_1V'W'-2WV F_1W''-2W^2F_1V''-3WV W'F'_1-2W^2V'F'_1
+{2V W^2F''_1} \right] \right. \nn
& \left. -4rWV \left[ F_1W'-W F'_1 \right]-{4W^2F_1 \left[ 1-V \right]}-8W r^2 {\eta'}^2 \right\}\,, \\
\label{ffrr}
0=& \frac{1}{8r^2W^2} \left\{r^2 \left[ V F_1{W'}^2-3W F_1V'W'-2WV F_1W''-2W^2F_1V''+WV W'F'_1-2W^2V'F'_1
 -6V W^2F''_1 \right] \right. \nn
& \left. +4rWV \left[ F_1W'+WF'_1 \right] -4W^2F_1 \left[1-V\right]-8W r^2 {\eta'}^2 \right\} \,,\\
\label{ffthth}
0=& \frac{1}{8r^2W^2} \left\{r^2 \left[ 3W F_1V'W'+2WV F_1W''+2W^2F_1V''-V F_1{W'}^2+WV W'F'_1+2W^2V'F'_1
+2V W^2F''_1 \right] \right. \nonumber\\
& \left. -4rW^2V F'_1+4W^2F_1 \left[ 1-V \right]+8W r^2 {\eta '}^2 \right\}\,.
\end{align}
Other components of the $f(R)$ equation vanish.

 {The trace of the field equation $f(R)$, given by Eq.~(\ref{f3}), takes the form:}
\begin{align}
0=\frac{1}{2r^2W^2} &\left\{ r^2 \left[6W^2V'F'_1-3W F_1V'W'-2WV F_1W''-2W^2F_1V''+V F_1{W'}^2+3WV W'F'_1
+6V W^2F''_1 \right] \right. \nonumber\\
& \left. +4rW \left[ 3WV F'_1-F_1V W'-2F_1WV' \right]+4W^2F_1 \left[1-V \right]-4r^2W^2f(r) \right\} \,,
\label{feq}
\end{align}
where $W(r)=\frac{U(r)}{V(r)}$ and $F_1\equiv F_1(r)=\frac{df(R(r))}{dR(r)}$, $F'_1=\frac{dF_1(r)}{dr}$,
$F''_1=\frac{d^2F_1(r)}{dr^2}$, $F'''_1=\frac{d^3F_1(r)}{dr^3}$.
{The Maxwell field equations have  the following  form:}
\begin{equation}
\label{feqc}
\frac{\eta ' \left[ rW'-4W \right]-2rW \eta ''}{2rW^2}=0\,,
\end{equation}
where $\eta$ is the component of the electric field, i.e., $\eta_\alpha=(\eta(r),0,0,0)$.
Using Eqs.~(\ref{fftt}) and (\ref{ffrr}), i.e., (\ref{fftt}) minus (\ref{ffrr}), we obtain
\begin{equation}
\label{E4n}
0 = r^2 \left[ - 4WV W'F'_1 + 8VW^2 F''_1 \right] - 8r WW'V F_1 \, .
\end{equation}
Moreover, Eqs.~(\ref{fftt}) and (\ref{ffthth}), i.e., (\ref{fftt}) plus (\ref{ffthth}) give,
\begin{equation}
\label{E6n}
0 = - 2 r^2 WVW'F'_1 + 4r^2VW^2 F''_1 - 4r WW' VF_1 \,.
\end{equation}
Eqs.~(\ref{E4n}) and (\ref{E6n}) coincide with each other.
Therefore we derive two independent equations from Eqs.~(\ref{fftt}), (\ref{ffrr}), and (\ref{ffthth}).
From  the above calculations, it is easy to prove that Eq.~(\ref{fftt}) is equal
Eq.~(\ref{ffrr}) with minus sign and equal minus two times Eq.~(\ref{ffthth}).
Hence, Equations~(\ref{fftt}) and (\ref{E6n}) are independent equations,
{which include all the possible solutions.}
Now we have four unknown functions $V$, $W$, $\eta $, and {$F_1$}, which is the reason why we are not able to determine
one function\footnote{Note that Eq.~(\ref{feqc}) can determine the unknown $\eta$.}.
{In order to obtain concrete and specific solutions and investigate the physical properties of the solutions in order to show
that Eqs.~(\ref{fftt}) and (\ref{E6n}) include physically reasonable and natural solutions, we make an assumption of the form
on {$F_1$} in the following.}

In our previous studies, we showed that when $W=1$ and providing $V\neq 0$,
we get  from Eq.~(\ref{E4n})
\begin{equation}
\label{ES2n}
F''_1=0 \, , \quad \mbox{that leads,} \quad
F_1=f_2 + f_3 r \, .
\end{equation}
From  Eqs.~(\ref{ES2n}) and (\ref{fftt}), we obtain
\begin{align}
\label{ES3n}
0=& r^2\left[ - 2 F_1V'' - 2 V'F'_1 \right] + 4r V F'_1 - 4F_1 \left[1 - V \right]-8 r^2 {\eta'}^2 \nn
=& - 2 r^2 \left( f_2 + f_3 r \right) V'' -2 r^2 f_3 V' + 4 \left( f_2 + 2 f_3 r \right) V
 - 4 \left( f_2 + f_3 r \right)-8r^2 {\eta'}^2 \, .
\end{align}
Assuming $f_3=0$, Eq.~(\ref{ES3n}) gives
\begin{equation}
\label{ES4n}
0= f_2 [r^2 V''- 2 V + 2]+4 r^2 {\eta'}^2 \, .
\end{equation}
From Eq.~(\ref{ES4n}) after using Eq.~(\ref{feqc}), we obtain the following solution
\begin{equation}
\label{ES5n}
V= 1 + \frac{V_0}{r} + V_1 r^2-\frac{V_2}{f_2 r^2}\,, \quad {\eta =\frac{\sqrt{V_2}}{r}}\, ,
\end{equation}
where $V_0$, $V_1$, and $V_2$ are integration constants.
Equation~(\ref{ES5n}) is the well-known Reissner-Nordstr\"om-(anti-)de Sitter space-time.

We studied the case $f_2=0$ in which Eq.~(\ref{ES3n}) gives,
\begin{equation}
\label{ES6n}
0 =- 2 r^2 f_3 r V'' -2 r^2 f_3 V' + 8 f_3 r V
 - 4 f_3 r -8r^2 {\eta'}^2\, .
\end{equation}
The solution of Eq.~(\ref{ES6n}) together with Eq.~(\ref{feqc}) have the following solution
\begin{equation}
\label{ES7n}
V= \frac{1}{2} + \frac{{\tilde V}_0 r^2}{20} + \frac{{\tilde V}_1}{ r^{2}}{- \frac{{\tilde V}_2}{r^{3}}} \,,
\quad \eta =\frac{\sqrt{5f_3 \tilde V_2}}{2 r}\, .
\end{equation}
{Here ${\tilde V}_0$, ${\tilde V}_1$, and ${\tilde V}_2$ are integration constants, again.}
Equation (\ref{ES7n}) corresponds to the solution derived in \cite{Nashed:2019tuk,Elizalde:2020icc}.

In \cite{Nashed:2019tuk,Elizalde:2020icc}, we also solved the system of differential equations (\ref{feq})
by assuming $F_1=1+\frac{c}{r^2}$.
In the present paper, we solve such a system by assuming
\begin{equation}
\label{const}
F_1=1-\frac{F_0-\left(n-3\right)}{r^n}\,,
\end{equation}
where $n$ can take any value and $F_0$ is a constant.
{In the nominator of the second term in the r.h.s. of (\ref{const}), $(n-3)$ can be absorbed
into the redefinition of $F_0$ but we use the form of (\ref{const}) for later convenience.}

{
Using Eq.~(\ref{const}) in the system (\ref{fftt}), (\ref{ffrr}), and (\ref{ffthth}), we obtain
\begin{align}
0=& r^2\left( {r}^{n}-F_0+n-3 \right) \left[ 2W^2V'' +2V W  W''  -V W'^{2}+3W V' W' \right]
+ 4W^2\left(1 -V \right) r^n +8Wr^{2+n}\eta'^2\nonumber\\
& +W \left[ 4r^n+(3n-4)\left( F_0-n+3\right) \right]VrW'
+2W^2\left( F_0-n+3\right)\left[ nr V'- \left( 2- \left( {n}^{2}-\,n+2 \right) V  \right)  \right]\,,
\label{E11n} \\
& r^2\left( r^n -F_0+n-3 \right) \left[ 2W^{2}V'' +2V W  W'' -V W'^2 +3W V' W' \right]
+ 4W^2\left(1 -V \right) r^n +8Wr^{2+n}\eta'^2 \nonumber\\
&-W \left[ 4r^n+(n-4)\left( F_0-n+3\right) \right]VrW'
+2\,W^2\left( F_0-n+3\right)\left[ nr V' - \left\{ 2+ \left( 3{n}^{2}+5\,n-2 \right) V  \right\}  \right]\,,
\label{E2n}\\
0=& r^2\left( r^n -F_0+n-3 \right) \left[ 2\,W^2 V'' +2V W  W''  -V W'^{2}+3W V' W' \right]
+ 4W^2\left(1 -V \right) r^n \nonumber\\
&+8Wr^{2+n}\eta'^2+W \left( F_0-n+3\right) \left[nVrW'+2\,W \left\{ nr  V' - \left( 2+ \left( n^2
+3 n-2 \right) V \right) \right\} \right]\,.
\label{E3n}
\end{align}}
We then solve the above system of differential equations in cases ${\eta}=0$
and ${\eta}\neq 0$.

\subsection{The case of ${\eta}=0$}

The analytic solution of the above system takes the following form,
\begin{align}
\label{ass1}
V(r)=& \frac{r^{\frac{2 \left( n^2+2n-2 \right)}{n-2}}}{X_1^{\frac{2 \left( n+1 \right)}{n-2}}} \left[ c_2
+ \mathlarger{\mathlarger{\int}} \frac{X_1^{\frac{n+1}{n-2}}}{r^{\frac{7n-8}{n-2}} X_2}dr
\left(c_1-2 \mathlarger{\mathlarger{\int}}
\frac{X_1^{\frac{n+1}{n-2}}X_2}{c_3r^{\frac{n(2n-1)}{n-2}}}dr \right)
+2 \mathlarger{\mathlarger{\mathlarger{\int}}} \frac{X_2 X_1^{\frac{n+1}{n-2}} {\mathlarger{\int}
\frac{X_1^{\frac{n+1}{n-2}}}{r^{\frac{7n-8}{n-2}} X_2}dr}}{c_3r^{\frac{n(2n-1)}{n-2}}}dr \right]\,,\nonumber\\
W(r)=& c_3r^{\frac{2n(n+1)}{2-n}}X_1^{\frac{2(n+1)}{n-2}}\,,\quad U(r)=W(r)V(r)\,, \quad F_1=1-\frac{F_0-\left(n-3\right)}{r^n}\,,
\end{align}
where $X_1(r)= \frac{F_0 \left( n-2 \right)}{2}-\frac{(n-2)(n-3)}{2}+r^n$ and $X_2(r)= F_0-n+3-r^n$.
{ Because there appear the fractional powers of $X_1$ in the expressions of $V$ and $W$ (and $U$),
if we require $X_1>0$ for any $r$ to avoid that the complex number appears in $V$ and $W$,
we find $F_0 - (n-3)\geq 0$. }

As Eq.~(\ref{ass1}) shows, the case of $n=2$ is not allowed.
However this case was studied in \cite{Nashed:2020mnp} and we obtained
\begin{align}
\label{assp}
V(r)=&\frac{\e^{\frac{3c_1}{2r^2}}}{r}\left\{\mathrm{H}c_2+\mathrm{H}_1r^3c_3
+2\mathrm{H}_1 r^3 \mathop{\mathlarger{\int}} \frac{\e^{-\frac{3c_1}{2r^2}}\mathrm{H}}{r \left[ \left( 2c_1\mathrm{H}_2
 - 3r^2\mathrm{H} \right) \mathrm{H}_1-2c_1\mathrm{H}\mathrm{H}_3 \right]}dr \right. \nn
& \left. \qquad \qquad -2\mathrm{H} \mathop{\mathlarger{\int}}
\frac{\e^{-\frac{3c_1}{2r^2}}r^2 \mathrm{H}_1}{ \left( 2c_1\mathrm{H}_2
 - 3r^2\mathrm{H} \right)\mathrm{H}_1-2c_1\mathrm{H}\mathrm{H}_3}dr \right\}\,,\nonumber\\
U(r)=& \e^{^{\frac{3c_1}{2r^2}}}\,, \quad W(r)=N(r)B(r)\,, \quad F_1=1+\frac{c_1}{r^2}\,,
\end{align}
where $\mathrm{H}=\mathrm{HeunC} \left( \frac{3}{2},\frac{3}{2},0,\frac{3}{8},\frac{9}{8},-\frac{c_1}{r^2} \right)$,
$\mathrm{H}_1=\mathrm{HeunC} \left( \frac{3}{2},-\frac{3}{2},0,\frac{3}{8},\frac{9}{8},-\frac{c_1}{r^2} \right)$,
$\mathrm{H}_2=\mathrm{HeunCPrime} \left( \frac{3}{2},\frac{3}{2},0,\frac{3}{8},\frac{9}{8},-\frac{c_1}{r^2} \right)$, and
$\mathrm{H}_3=\mathrm{HeunCPrime} \left( \frac{3}{2},-\frac{3}{2},0,\frac{3}{8},\frac{9}{8},-\frac{c_1}{r^2} \right)$\footnote{
The special function $\mathrm{HeunC}$ is defined as the solution of the Heun Confluent equation, that has the form
\[
X''(r)-\frac{1+\beta-(\alpha-\beta-\gamma-2)r-r^2\alpha}{r(r-1)}X'(r)
 -\frac{\alpha(1+\beta)-\gamma-2\eta-(1+\gamma)\beta-r(2\delta+[2+\gamma+\beta])}{2r(r-1)}X(r)=0\,.
\]
The  above differential equation has the following solution:  $\mathrm{HeunC} \left( \alpha,\beta,\gamma,\delta,\eta,r \right)$.
Interested readers can check \cite{RONVEAUX2003177,MAIER2005171} for more details.
The special function $\mathrm{HeunCPrime}$ is defined as the derivative of the Heun Confluent function.}.
Therefore, as Eqs.~(\ref{ass1}) and (\ref{assp}) show the case of $n=2$ is defined by the function $\mathrm{HeunC}$
while the case of $n \neq 2$ is defined by (\ref{ass1}). We will discuss this case in detail below.

Using Eq.~(\ref{ass1}) in (\ref{feq}), we derive the form of $f(r)$ as follows,
\begin{align}
\label{ass11}
f(r) = & \frac{12}{r^{n+2}X_1^{\frac{3n}{n-2}}X_2} \left\{ r^\frac{2 \left( n^2+2n-2 \right)}{n-2} \left[
f_1^2(n^2-1)-\frac{(n^2-4)r^n}{2}f_1-r^{2n} \right] X_2 \left\{ {\mathop{\mathlarger{\mathlarger{\mathlarger{\int}}}}
\frac{X_1^{\frac{n+1}{n-2}}X_2 {\mathop{\mathlarger{\int}} \frac{X_1^{\frac{n+1}{n-2}}}{r^{\frac{7n-8}{n-2}}
X_2}dr}}{r^{\frac{n(2n-1)}{n-2}}}dr} \right. \right. \nonumber\\
& \left. -{ \mathop{\mathlarger{\mathlarger{\int}}}\frac{X_1^{\frac{n+1}{n-2}}}{r^{\frac{7n-8}{n-2}}
X_2}dr}\left[{\mathlarger{\mathlarger{\int}}\frac{X_1^{\frac{n+1}{n-2}}
X_2}{r^{\frac{n(2n-1)}{n-2}}}dr}-\frac{c_1}{2}\right] \right\} \nn
&  -r^{\frac{2 \left( n^2-n+1 \right)}{n-2}} \left( \left(F_0-n+3\right) \left( n+1 \right) -r^n\right)X_1^{\frac{(2n-1)}{n-2}}
{\mathop{\mathlarger{\mathlarger{ \int}}}\frac{X_1{}^{\frac{n+1}{n-2}}X_2}{3r^{\frac{n(2n-1)}{n-2}}
}dr} -\frac{1}{6}X_1^{\frac{3n}{n-2}}X_2{}^2 \nonumber\\
& +\frac{1}{2}c_2r^\frac{2 \left( n^2+2n-2 \right)}{n-2}X_2
\left[ \left\{f_1 \left( n^2-1 \right) - \frac{ \left( n^2-4 \right) r^n}{2}\right\}f_1
 -r^{2n} \right] \nn
 & \left. +\frac{1}{6}c_1r^{\frac{2 \left( n^2-n+1 \right)}{n-2}}X_1^{\frac{2n-1}{n-2}}\left(f_1(n+1)-r^n\right) \right\}\,,
\end{align}
where $f_1= F_0-n+3$.
Using Eq.~(\ref{ass11}) in Eq.~(\ref{Ricci}), we obtain the Ricci scalar in the form
\begin{align}
\label{sol11}
R =& -\frac{12}{r^2X_1^{\frac{3n}{n-2}}X_2^2} \left\{ r^\frac{2 \left( n^2+2n-2 \right)}{n-2}
\left\{ \frac{f_1(n-1)(n+4)}{2}+2r^n \right\} X_2^2 \left\{ {\mathop{\mathlarger{\mathlarger{\mathlarger{\int}}}}
\frac{X_2X_1^{\frac{n+1}{n-2}}{\mathop{\mathlarger{\int}}\frac{X_1^{\frac{n+1}{n-2}}}{r^{\frac{7n-8}{n-2}}
X_2}dr}}{r^{\frac{n(2n-1)}{n-2}}}dr} \right. \right. \nonumber\\
& \left. - \left[ \mathlarger{\mathlarger{\int}}\frac{X_1^{\frac{n+1}{n-2}} X_2}{r^{\frac{n(2n-1)}{n-2}}} dr
 -\frac{c_1}{2} \right] {\mathop{\mathlarger{\mathlarger{\int}}} \frac{X_1^{\frac{n+1}{n-2}}}{r^{\frac{7n-8}{n-2}} X_2}dr} \right\}
 -r^{\frac{2\left( n^2-n+1 \right)}{n-2}}\left(\frac{f_1(n+4)}{2}-2r^n\right)X_1^{\frac{(2n-1)}{n-2}}
{\mathop{\mathlarger{\mathlarger{ \int}}}\frac{X_1^{\frac{n+1}{n-2}}X_2}{3r^{\frac{n(2n-1)}{n-2}}
}dr}\nonumber\\
& -X_2^2\left\{\frac{1}{3}X_1^{\frac{3n}{n-2}}-\frac{1}{2}c_2r^\frac{2\left( n^2+2n-2 \right)}{n-2}
\left[\frac{f_1(n-1)(n+4)}{2}+2r^{n}\right]\right\} \nn
& \left. +\frac{1}{6}c_1r^{\frac{2\left( n^2-n+1 \right)}{n-2}}X_1^{\frac{2n-1}{n-2}}\left(\frac{f_1(n+4)}{2}-2r^n\right)\right\}\,.
\end{align}

\subsection{The case of ${\eta}\neq 0$}

Now we are going to find an analytic solution of the system (\ref{E11n}), (\ref{E2n}), and (\ref{E3n})
in the case of $\eta\neq 0$ and obtain
\begin{align}
\label{ass1c}
V(r)=&\frac{r^{\frac{2\left( n^2+2n-2 \right)}{n-2}}}{X_1^{\frac{2(n+1)}{n-2}}} \left[
c_2+ {\mathop{\mathlarger{\mathlarger{\int}}}\frac{X_1^{\frac{n+1}{n-2}}}{r^{\frac{7n-8}{n-2}}
X_2}dr}\left(c_1-2{\mathop{\mathlarger{\int}} \frac{X_1^{\frac{n+1}{n-2}}X_3}{c_3r^{\frac{n(2n-1)}{n-2}}}dr}\right)
+2{\mathop{\mathlarger{\mathlarger{\mathlarger{\int}}}}\frac{X_3 X_1^{\frac{n+1}{n-2}} {\mathop{
\mathlarger{\int}}\frac{X_1^{\frac{n+1}{n-2}}}{r^{\frac{7n-8}{n-2}}
X_2}dr}}{c_3r^{\frac{n(2n-1)}{n-2}}}dr}\right]\,,\nonumber\\
W(r)=&c_3r^{\frac{2n(n+1)}{2-n}}X_1^{\frac{2(n+1)}{n-2}}\,,\quad U(r)=W(r)V(r)\,, \quad 
F_1=
1-\frac{F_0-\left(n-3\right)}{r^n}\,, \quad
\eta(r)=c_4+c_5{\mathop{\mathlarger{\mathlarger{\int}}}\frac{X_1^{\frac{n+1}{n-2}}}{r^{\frac{(n-1)(n+4)}{n-2}}}dr}\,,
\end{align}
where $X_3= c_3X_2-2c_5^2r^{n-2}$.
Equation (\ref{ass1c}) reduces to (\ref{ass1}) when $\eta=0$.
As Eq.~(\ref{ass1c}) shows the case $n=2$ is not allowed, again.
However, this case was also studied in \cite{Nashed:2020tbp}
where we obtained
\begin{align}
\label{ass1cp}
V(r)=& \frac{\e^{\frac{3a_1}{2r^2}}}{r}\left\{\mathrm{H}a_2
+\mathrm{H}_1r^3a_3+2\mathrm{H}_1r^3 \mathlarger{\int} \frac{\e^{-\frac{3a_1}{2r^2}}
\mathrm{H} \left[ r^2+13a_1 \right]}{r(r^2+a_1) \left[ \left(2a_1\mathrm{H}_2
 - 3r^2\mathrm{H} \right) \mathrm{H}_1-2a_1\mathrm{H}\mathrm{H}_3 \right]}dr
\right. \nonumber\\
& \left. -2\mathrm{H} \mathlarger{\int} \frac{\e^{-\frac{3a_1}{2r^2}}r^2 \mathrm{H}_1 \left[r^2+13a_1 \right]}
{\left( r^2+a_1 \right) \left[ \left( 2a_1\mathrm{H}_2
 - 3r^2\mathrm{H} \right)\mathrm{H}_1-2a_1\mathrm{H}\mathrm{H}_3 \right]}dr
\right\}\,,\nonumber\\
W(r)=&\e^{- \frac{3a_1}{r^2}}\,,\qquad U(r)=V(r)W(r)\,,\quad
\eta=a_0+\sqrt{\pi}\,\mathrm{erf} \left[\frac{\sqrt{6a_1}}{2r}\right]\,.
\end{align}
Using Eq.~(\ref{ass1c}) in the trace equation given by Eq.~(\ref{feq}),
we obtain the following form of $f(r)$ as\footnote{{ The functions $f(r)$ given by Eqs.~(\ref{ass11}) and (\ref{frc})
are constrained by the condition that their first derivative
$f_R=\frac{\partial f(R)}{\partial R}=\frac{\partial f(r)}{\partial r}\times \frac{\partial r}{\partial R}=F(r)=1+\frac{c_1}{r^2}$}.}:
\begin{align}
\label{frc}
f(r)=&\frac{12}{c_3r^{n+2}X_1^{\frac{3n}{n-2}}X_2} \left\{ r^\frac{2\left( n^2+2n-2 \right)}{n-2}\left[
f_1^2(n^2-1)-\frac{(n^2-4)r^n}{2}f_1-r^{2n} \right] X_2 \left\{{\mathop{\mathlarger{\mathlarger{\mathlarger{\int}}}}
\frac{X_1^{\frac{n+1}{n-2}}X_3{\mathop{
\mathlarger{\mathlarger{\int}}}\frac{X_1^{\frac{n+1}{n-2}}}{r^{\frac{7n-8}{n-2}}
X_2}dr}}{r^{\frac{n(2n-1)}{n-2}}}dr} \right. \right. \nonumber\\
& \left. - {\mathop{\mathlarger{\mathlarger{\int}}}\frac{X_1^{\frac{n+1}{n-2}}}{r^{\frac{7n-8}{n-2}}
X_2}dr} \left[ \mathop{
\mathlarger{\mathlarger{\int}}}\frac{X_1^{\frac{n+1}{n-2}}
 X_3}{r^{\frac{n(2n-1)}{n-2}}}dr - \frac{c_3c_1}{2}\right] \right\} \nn
& -r^{\frac{2\left( n^2-n+1 \right)}{n-2}}\times \left(\left(F_0-n+3\right) (n+1)-r^n\right)X_1^{\frac{(2n-1)}{n-2}}
{\mathop{\mathlarger{\mathlarger{ \int}}}\frac{X_1{}^{\frac{n+1}{n-2}}X_3}{3r^{\frac{n(2n-1)}{n-2}}
}dr}\nonumber\\
& -\frac{1}{6}X_1^{\frac{3n}{n-2}}X_2X_4+\frac{1}{2}c_2r^\frac{2\left( n^2+2n-2 \right)}{n-2}X_2
\left[ \left\{ f_1 \left(n^2-1 \right) - \frac{\left( n^2-4 \right)r^n}{2}\right\}f_1-r^{2n}\right] \nn
& \left. +\frac{1}{6}c_1c_3r^{\frac{2\left( n^2-n+1 \right)}{n-2}}X_1^{\frac{2n-1}{n-2}}\left(f_1(n+1)-r^n\right)\right\}\,,
%
\end{align}
where $X_4= c_3X_2-c_5^2r^{n-2}$.
Using Eq.~(\ref{ass1c}) in Eq.~(\ref{Ricci}), we obtain the Ricci scalar in the following form
\begin{align}
\label{sol1c}
R =& -\frac{12}{c_3r^2X_1^{\frac{3n}{n-2}}X_2^2}\left\{ r^\frac{2\left( n^2+2n-2 \right)}{n-2}
\left\{ \frac{f_1(n-1)(n+4)}{2} + 2r^n \right\} X_2^2 \left\{ \mathlarger{\mathlarger{\mathlarger{\int}}}
\frac{X_3X_1^{\frac{n+1}{n-2}}{\mathlarger{\mathlarger{\int}}
\frac{X_1^{\frac{n+1}{n-2}}}{r^{\frac{7n-8}{n-2}} X_2}dr}}{r^{\frac{n(2n-1)}{n-2}}}dr \right. \right. \nonumber\\
& \left. - \left[\mathlarger{\mathlarger{\int}}\frac{X_1^{\frac{n+1}{n-2}}
X_3}{r^{\frac{n(2n-1)}{n-2}}}dr-\frac{c_1}{2}\right] {\mathop{\mathlarger{\mathlarger{ \int}}}
\frac{X_1^{\frac{n+1}{n-2}}}{r^{\frac{7n-8}{n-2}}
X_2}dr}\right\} -r^{\frac{2\left( n^2-n+1 \right)}{n-2}}\left(\frac{f_1(n+4)}{2}-2r^n\right)X_1^{\frac{(2n-1)}{n-2}}
{\mathop{\mathlarger{\mathlarger{ \int}}}\frac{X_1^{\frac{n+1}{n-2}}X_3}{3r^{\frac{n(2n-1)}{n-2}}
}dr}\nonumber\\
& -X_2\left\{ \frac{1}{3}X_4\,X_1^{\frac{3n}{n-2}}-\frac{1}{2}c_2c_3r^\frac{2\left( n^2+2n-2 \right)}{n-2}X_2
\left[ \frac{f_1(n-1)(n+4)}{2}+2r^{n} \right] \right\} \nn
& \left. +\frac{1}{6}c_1c_3r^{\frac{2\left( n^2-n+1 \right)}{n-2}}X_1^{\frac{2n-1}{n-2}}\left(\frac{f_1(n+4)}{2}-2r^n\right)\right\}\,.
\end{align}
In the following section,  we will study the physics beyond the BH solutions with/without charge given by Eq.~(\ref{ass1}) and (\ref{ass1c})

\section{Inherent physics of the BH solutions (\ref{ass1}) and (\ref{ass1c})}\label{S4}

An important detail to emphasize is that when $F_0=n-3$, we recover the GR BH, as Eqs.~(\ref{ass1}) and (\ref{ass1c}) show.
Now we extract the inherent physics of the BH solutions (\ref{ass1}) and (\ref{ass1c}).
We therefore concentrate on the case $n=4$ and write the metric potentials of the neutral BH solution (\ref{ass1}) as
\begin{align}
\label{ass2n}
V(r)=& \frac{c_2r^{22}}{ \left( F_0 -1+r^4 \right)^5} + \frac{r^{22}}{\left( F_0 -1+r^4 \right)^5}
\left\{ c_1\mathlarger{\mathlarger{\int}} \frac{\left(F_0 -1+r^4 \right)^{\frac{5}{2}}}{r^{10}
\left( F_0 - 1 - r^4 \right)^5}dr \right. \nn
& +2{\mathop{\mathlarger{\mathlarger{\mathlarger{\int}}}}\frac{ \left( F_0-1-r^4 \right){\left[\mathop{
\mathlarger{\mathlarger{\int}}}\frac{\left( F_0-1+r^4 \right)^{\frac{5}{2}}}
{r^{10} \left( F_0-1-r^4 \right)}dr\right]}\left( F_0-1+r^4 \right)^{\frac{5}{2}}}{r^{14}}dr}
\nonumber\\
& \left. -2{\mathop{\mathlarger{\mathlarger{\mathlarger{\int}}}}
\frac{\left( F_0 -1+r^4 \right)^{\frac{5}{2}} \left( F_0-1-r^4 \right)}{r^{14}}dr}
{\mathop{\mathlarger{\mathlarger{\int}}}\frac{\left( F_0 -1 + r^4 \right)^{\frac{5}{2}}}{r^{10}
\left( F_0-1-r^4 \right)}dr} \right\} \, , \nonumber\\
W(r)=&\frac{c_3 \left( F_0 -1 + r^4 \right)^5}{r^{20}}\,,\quad
U(r)=W(r)V(r)\,, \quad F_1=1-\frac{F_0-1}{r^4}\, .
\end{align}
{ We should note that $F_0-1$ should not be negative in order that $\left( F_0 -1 + r^4 \right)^{\frac{5}{2}}$ should be a real
number for any value of $r$. }
The asymptotic forms of the metric potentials { when $r$ is large} given by Eq.~(\ref{ass2n}) take the following forms
\begin{align}
\label{ass3n}
V(r)=& r^2c_2+1+\frac{c_1}{3r} - \left( F_0 - 1 \right)
\left[\frac{5c_2}{r^2}+\frac{4}{r^4}+\frac{7c_1}{6r^5}-\frac{15c_1 \left( F_0 - 1 \right)}{r^6}+\cdots\right]\,,\nonumber\\
U(r)=& r^2c_2+1+\frac{c_1}{3r} + \left( F_0 - 1 \right)
\left[\frac{1}{r^4}+\frac{c_1}{2r^5}+\frac{26 \left( F_0 - 1 \right)}{35r^8}+\cdots \right]\,,
\end{align}
where we have chosen $c_3=1$. Using Eq.~(\ref{ass3n}) in Eq.~(\ref{met12}), we
obtain the line element in the form
\begin{align}
\label{metafn}
ds^2 \approx & -\left\{r^2\Lambda_\mathrm{eff}+1-\frac{2m}{r}+\left( F_0 - 1 \right)
\left[\frac{1}{r^4}-\frac{3m}{r^5}+\frac{26\left( F_0 - 1 \right)}{35r^8}\right]\right\} dt^2\nonumber\\
&+\frac{dr^2}{r^2\Lambda_\mathrm{eff}+1-\frac{2m}{r}-\left( F_0 - 1 \right)
\left[\frac{5\Lambda_\mathrm{eff}}{r^2}+\frac{4}{r^4}-\frac{7m}{r^5}
+ \frac{90m\left( F_0 - 1 \right)}{r^6}\right]} + d\Omega^2
\,,
\end{align}
where 
$m=-\frac{c_1}{6}$ and $c_2=\Lambda_\mathrm{eff}$.

Equation~(\ref{metafn}) shows that the line element expresses the asymptotic anti-de Sitter (A)dS space-time and
is not identical with the Schwarzschild space-time due to the contribution of the extra terms of the higher-order curvature
of the $f(R)$ gravity.
Equation~(\ref{metafn}) ensures what we have stated in the introduction, that is, in the $f(R)$ gravity, one can derive a space-time
that is different from the Schwarzschild-(A)dS one and when  $F_0=1$, i.e., $F_1=1$, we recover the Schwarzschild (A)dS
metric \cite{Misner:1974qy} as usual.
In conclusion, at a higher-order curvature, we can obtain a neutral space-time that is unlike the Schwarzschild solution
and coincides with the Schwarzschild (A)dS at a lower order of $f(R)=  R+ \mathrm{constant}$.

Now we use Eq.~(\ref{ass3n}) in Eq.~(\ref{Ricci}) and obtain\footnote{{The asymptote of Ricci scalar is given as
\[
R(r)=-12\Lambda_\mathrm{eff}-\frac{12\left( F_0 - 1 \right)}{r^6}\left[3+\frac{8m}{r}
 -\frac{15\left( F_0 - 1 \right)\Lambda_\mathrm{eff}}{r^2}+O\left(\frac{1}{r^4}\right)\cdots\right]\,.
\]}}
\begin{align}
\label{R1}
R(r)\approx& -12\Lambda_\mathrm{eff}-\frac{36\left( F_0 - 1 \right)}{r^6}\ \Rightarrow\
r(R)=\frac{6^{1/3}\left[ \left( R+12\Lambda_\mathrm{eff} \right)^5\left( 1 - F_0 \right) \right]^{1/6}}
{R+12\Lambda_\mathrm{eff}}\,,
\end{align}
where we have omitted the other terms in Ricci scalar to be able to write the radial coordinate $r$ as a function of Ricci scalar, i.e., $r(R)$.
 From Eq.~(\ref{R1}), we can clearly see that when $F_0=1$, we obtain a constant value of the Ricci scalar
because when $F_1=1$ and $f_R=\mathrm{const}$.
The asymptote of $f(r)$ given by Eq.~(\ref{ass11}) becomes
\begin{eqnarray}
\label{fR1}
f({r})\approx -6\Lambda_\mathrm{eff}-\frac{36\left( F_0 - 1 \right)}{r^6}
+\frac{96m\left( F_0 - 1 \right)}{r^7}\cdots\,.
\end{eqnarray}
Using Eq.~(\ref{R1}) in (\ref{fR1}), we obtain
\begin{eqnarray}
\label{fR2}
f(R)\approx 6\Lambda_\mathrm{eff}+ R -\frac{12m6^{2/3} R ^{7/6}}{27\left( 1 - F_0 \right)^{1/6}}-\frac{168m\Lambda_\mathrm{eff}6^{2/3} R ^{1/6}}{27\left( 1 - F_0 \right)^{1/6}}
-\frac{168m\,\Lambda_\mathrm{eff}{}^2\,6^{2/3}}{27\left( 1 - F_0 \right)^{1/6} R ^{5/6}}
\cdots\,.
\end{eqnarray}
{
 Equation (\ref{fR2}) shows that $f(R)$ include the term with the higher positive power of $R$ than the Einstein--Hilbert term,
which may dominate when $R$ is large,
and also the term with negative power, which may dominate when $R$ is small. }

The expression of Eq.~(\ref{fR2}) might look strange because there is a divergence in the limit of $F_0\to 1$,
which should correspond to the standard Einstein-Hilbert action with a cosmological constant.
This occurs because we have used an expansion assuming that $r$ is large in Eq.~(\ref{fR1}), but after that,
if we consider the limit of $F_0\to 1$, the scalar curvature $R$ becomes a constant, which means
that $R$ is independent of $r$ and therefore we cannot solve $r$ with respect to $R$.
Then it is natural if the expression (\ref{fR2}), which is obtained by combining (\ref{R1}) and (\ref{fR1}),
becomes singular in the limit that of $F_0 \to 1$.
The singularity is rather an artificial one coming from the non-commutability of the two limits
that $r \to \infty$ and $F_0 \to 1$ but the behavior makes the situations ambiguous and often uncontrollable.
In order to avoid this difficulty, we also use the numerical calculations.

Using Eq.~(\ref{ass3n}), we obtain the { curvature} invariants of solution (\ref{ass1}) as
\begin{align}
\label{invn}
R_{\mu \nu \rho \sigma}  R ^{\mu \nu \rho \sigma}=& R _{\mu \nu }  R ^{\mu \nu }= 24\Lambda_\mathrm{eff}{}^2
+\frac{144\Lambda_\mathrm{eff}\left( F_0 - 1 \right)+48m^2}{r^6}-\frac{384\Lambda_\mathrm{eff}\,m\left( F_0 - 1 \right)+48m^2}{r^7}-
\frac{120\Lambda_\mathrm{eff}^2\,\left( F_0 - 1 \right)^2}{r^8}
+\cdots \nonumber\\
R =& -12\Lambda_\mathrm{eff}-\frac{36\left( F_0 - 1 \right)}{r^6}+\frac{32m\left( F_0 - 1 \right)}{r^7}+\cdots\,,
\end{align}
where $\left(  R _{\mu \nu \rho \sigma}  R ^{\mu \nu \rho \sigma},
 R _{\mu \nu}  R ^{\mu \nu},  R  \right)$ are
the Kretschmann scalar, the Ricci tensor square and the Ricci scalar, respectively, and all of them have
a true singularity when $r=0$. An important detail to highlight is the fact that   $F_0$
is the main reason of the differentiation of the present study from the (A)dS Schwarzschild BH solution
of GR whose invariants behave as $\left(  R _{\mu \nu \rho \sigma}
R ^{\mu \nu \rho \sigma},  R _{\mu \nu}  R ^{\mu \nu},
R  \right) = \left( 24\Lambda^2+\frac{48M^2}{r^6},36\Lambda^2, 12\Lambda \right)$.
Equation (\ref{invn}) shows that the leading order of the scalars
$\left( R _{\mu \nu \rho \sigma}  R ^{\mu \nu \rho \sigma}, R _{\mu \nu} R ^{\mu \nu}, R  \right)$
is $\left(\frac{1}{r^\mathrm{6}}, \frac{1}{r^\mathrm{6}},\frac{1}{r^\mathrm{6}} \right)$ which coincides
with the from of the (A)dS Schwarzschild BH solution whose leading terms of the Kretschmann
is $\left( \frac{1}{r^\mathrm{6}}\right)$. Thus, Eq.~(\ref{invn})
shows that the singularity of the Kretschmann coincides
with the (A)dS Schwarzschild BH solution of GR.

Now we apply the same procedure that was used for the neutral case to the charged one.
The analytic solution (\ref{ass1c}) in the case $n=4$ takes the form
\begin{align}
\label{ass2c}
V(r)=& \frac{r^{22}}{c_3\left( F_0 - 1 + r^4 \right)^5}
\left\{ c_2c_3 + c_1c_3\mathlarger{\mathlarger{\int}}\frac{\left( F_0 - 1 + r^4 \right)^{\frac{5}{2}}}{r^{10}
\left( F_0 - 1 - r^4 \right)^5}dr
 - 2 \mathlarger{\mathlarger{\int}} \frac{\left( F_0 - 1 + r^4 \right)^{\frac{5}{2}}
\left[ c_3\left(F_0 -1-r^4\right)-2c_5^2r^2\right]}{r^{14}}dr \right. \nonumber\\
& \left. \times \mathlarger{\mathlarger{\int}} \frac{\left( F_0 - 1 + r^4 \right)^{\frac{5}{2}}}{r^{10}
\left( F_0 - 1 - r^4 \right)}dr + 2 \mathlarger{\mathlarger{\int}} \frac{\left[c_3\left(F_0 -1-r^4\right)-2c_5^2r^2\right]
\left( F_0 - 1 + r^4 \right)^{\frac{5}{2}}{\left[ \mathlarger{\int}
\frac{\left( F_0 - 1 + r^4 \right)^{\frac{5}{2}}}{r^{10}\left( F_0 - 1 - r^4 \right)}dr\right]}}{r^{14}}dr
\right. \, , \nonumber\\
W(r)=&\frac{c_3\left( F_0 - 1 + r^4 \right)^5}{r^{20}}\,,\quad U(r)=W(r)V(r)\,, \quad F_1=1-\frac{F_0-1}{r^4}\,, \quad
\eta(r)=c_4+c_5\int\frac{\left( F_0 - 1 + r^4 \right)^{\frac{5}{2}}}{r^{12}}dr\,.
\end{align}
Equation (\ref{ass2c}) reduces to Eq.~(\ref{ass2n}) when the constant $c_5=0$ which makes the electric charge
$\eta(r)=\mathrm{const}$.
To further examine this charged solution, we calculate its asymptotic form and obtain
\begin{align}
\label{mpab}
V(r)\thickapprox& r^2c_2+1+\frac{c_1}{3r}+\frac{c_5^2}{r^2}-\left( F_0 - 1 \right)
\left[\frac{5c_2}{r^2}+\frac{4}{r^4}+\frac{7c_1}{6r^5}
+ \frac{3[c_5^2-5c_1\left( F_0 - 1 \right)]}{r^6}+\cdots\right]\,,\nonumber\\
U(r)\thickapprox& r^2c_2+1+\frac{c_1}{3r}+\frac{c_5^2}{r^2}+\left( F_0 - 1 \right)
\left[\frac{1}{r^4}+\frac{c_1}{2r^5}+\frac{2c_5^2}{r^6}+\frac{26\left( F_0 - 1 \right)}{35r^8}+\cdots\right]\,,
\end{align}
where we use $c_3=1$.
Using Eq.~(\ref{mpab}) in Eq.~(\ref{met12}), we obtain the line element in the following form
\begin{align}
\label{metaf}
ds^2\thickapprox& -\left\{r^2\Lambda_\mathrm{eff}+1-\frac{2m}{r}+\frac{q^2}{r^2}+\left( F_0 - 1 \right)
\left[\frac{1}{r^4}-\frac{3m}{r^5}+\frac{2q^2}{r^6}+\frac{26\left( F_0 - 1 \right)}{35r^8}\right]\right\} dt^2\nonumber\\
&+\frac{dr^2}{r^2\Lambda_\mathrm{eff}+1-\frac{2m}{r}+\frac{q^2}{r^2}-\left( F_0 - 1 \right)
\left[\frac{5\Lambda_\mathrm{eff}}{r^2}+\frac{4}{r^4}-\frac{7m}{r^5}
+ \frac{3[q^2+30m\left( F_0 - 1 \right)]}{r^6}\right]} + d\Omega^2
\,,
\end{align}
where $q=c_5$, $\Lambda_\mathrm{eff}$ and $m$ have the same values given in the neutral case.

Equation (\ref{metaf}) shows that the line element is asymptotic to (A)dS and is
not equivalent to the Reissner-Nordstr\"om space-time in respect of the contribution of the
extra terms of the higher-order curvature of the $f(R)$ gravity. Equation (\ref{metaf})
shows clearly that in the $f(R)$ gravity, one can obtain a space-time that is different from
the Reissner-Nordstr\"om space-time and when the constant $F_0=1$, we recover
the Reissner-Nordstr\"om (A)dS metric \cite{Misner:1974qy}.
We can summarize  the results of this section by saying that at a higher order curvature, we can obtain a
 charged space-time that is unlike Reissner-Nordstr\"om space-time and reduces to
the Reissner-Nordstr\"om one (A)dS at a lower order of $f(R)=R$.

Using Eq.~(\ref{mpab}), we obtain the invariants of solution (\ref{ass1c}) as:
\begin{align}
\label{invc}
R _{\mu \nu \rho \sigma}  R ^{\mu \nu \rho \sigma}=& R _{\mu \nu }  R ^{\mu \nu }= 24{\Lambda_\mathrm{eff}}^2
+\frac{144\Lambda_\mathrm{eff}\left( F_0 - 1 \right)+48m^2}{r^6}
 -\frac{384\Lambda_\mathrm{eff}\,m\left( F_0 - 1 \right)+96mq^2}{r^7}\nonumber\\
&-\frac{120\Lambda_\mathrm{eff}^2\,\left( F_0 - 1 \right)^2+56q^2 \left[ q^2+30m\left( F_0 - 1 \right) \right]}{r^8}
+\cdots \, , \nonumber\\
R =& -12\Lambda_\mathrm{eff}-\frac{36\left( F_0 - 1 \right)}{r^6}+\frac{32m\left( F_0 - 1 \right)}{r^7}
+\left( F_0 - 1 \right)\frac{180\Lambda_\mathrm{eff}\,\left( F_0 - 1 \right)+60q^2}{r^8}+\cdots\,.
\end{align}
Equation (\ref{invc}) shows that all the invariants suffer
a true singularity when $r=0$. A substantial detail to highlight is that   $F_0$
is the origin that make the above results unlike the Reissner-Nordstr\"om (A)dS BH solution
of GR whose scalars behave as $\left(  R _{\mu \nu \rho \sigma}
 R ^{\mu \nu \rho \sigma},  R _{\mu \nu}  R ^{\mu \nu},
 R  \right) = \left( 24\Lambda^2+\frac{48M^2}{r^6},36\Lambda^2+\frac{4q^4}{r^8},
12\Lambda \right)$.
Equation (\ref{invc}) shows that the leading expression of the scalars $\left(
 R _{\mu \nu \rho \sigma}  R ^{\mu \nu \rho \sigma}, R _{\mu \nu}
 R ^{\mu \nu}, R  \right)$ is $\left(\frac{1}{r^\mathrm{6}},
\frac{1}{r^\mathrm{6}},\frac{1}{r^\mathrm{6}} \right)$ which does not coincide with the from of
the Reissner-Nordstr\"om (A)dS BH solution whose leading expressions of the Kretschmann and the Ricci tensor
squared are $\left( \frac{1}{r^\mathrm{6}},\frac{1}{r^\mathrm{8}} \right)$. Thus, Eq.~(\ref{invc})
shows that the singularity of the Ricci tensor squared is much milder
than that of the Reissner-Nordstr\"om (A)dS BH solution of GR.


\section{Stability of the BHs using geodesic deviation}\label{S6336b}


The geodesic equations take the form \cite{Misner:1974qy},
\begin{equation}
\label{ge3}
\frac{d^2 x^\alpha}{d\varepsilon^2}
+ \left\{ \begin{array}{c} \alpha \\ \beta \rho \end{array} \right\}
\frac{d x^\beta}{d\varepsilon} \frac{d x^\rho}{d\varepsilon}=0\, ,
\end{equation}
where $\varepsilon$ is the affine connection parameter. The
geodesic deviation equations take the form \cite{1992ier..book.....D,Nashed:2003ee},
\begin{equation}
\label{ged333}
\frac{d^2 \epsilon^\sigma}{d\varepsilon^2}
+ 2\left\{ \begin{array}{c} \sigma \\ \mu \nu \end{array} \right\}
\frac{d x^\mu}{d\varepsilon} \frac{d \epsilon^\nu}{d\varepsilon}
+ \left\{ \begin{array}{c} \sigma \\ \mu \nu \end{array} \right\}_{,\, \rho}
\frac{d x^\mu}{d\varepsilon} \frac{d x^\nu}{d\varepsilon}\epsilon^\rho=0\, ,
\end{equation}
with $\epsilon^\rho$ being the four-vector deviation. Plugging Eqs.~(\ref{ge3}) and (\ref{ged333})
into Eq.~(\ref{met12}), we obtain
\begin{equation}
\label{ges}
\frac{d^2 t}{d\varepsilon^2}=0\, , \qquad \frac{1}{2} U'(r) \left(
\frac{d t}{d\varepsilon}\right)^2 - r\left( \frac{d \phi}{d\varepsilon}\right)^2=0\, , \qquad
\frac{d^2 \theta}{d\varepsilon^2}=0 \, ,\qquad \frac{d^2 \phi}{d\varepsilon^2}=0\, ,
\end{equation}
and for the geodesic deviation the line-element (\ref{met12}) gives
\begin{align}
\label{ged11}
& \frac{d^2 \epsilon^1}{d\varepsilon^2} +V(r)U'(r) \frac{dt}{d\varepsilon}
\frac{d \epsilon^0}{d\varepsilon} -2r V(r) \frac{d \phi}{d\varepsilon}\frac{d \epsilon^3}{d\varepsilon}
+\left[ \frac{1}{2} \left( U'(r)V'(r)+V(r) U''(r)
\right)\left( \frac{dt}{d\varepsilon} \right)^2-\left(V(r)+rV'(r)
\right) \left( \frac{d\phi}{d\varepsilon}\right)^2 \right]\epsilon^1=0\, ,
\nonumber\\
& \frac{d^2 \epsilon^0}{d\varepsilon^2} + \frac{V'(r)}{V(r)} \frac{dt}{d\varepsilon}
\frac{d {\epsilon^1}}{d\varepsilon}=0\, ,\qquad \frac{d^2 \epsilon^2}{d\varepsilon^2}
+\left( \frac{d\phi}{d\varepsilon}\right)^2 \epsilon^2=0\, , \qquad
\frac{d^2 \epsilon^3}{d\varepsilon^2} + \frac{2}{r} \frac{d\phi}{d\varepsilon}
\frac{d \epsilon^1}{d\tau}=0\, ,
\end{align}
where $U(r)$ and $V(r)$ are given in Eq.~(\ref{ass3n}) or Eq.~(\ref{mpab}) and
$'$ is the  derivative with respect to the radial coordinate $r$.
From the condition of  a circular orbit, we can obtain
\begin{equation}
\label{so}
\theta= \frac{\pi}{2}\, , \qquad
\frac{d\theta}{d\varepsilon}=0\, , \qquad \frac{d r}{d\varepsilon}=0\, .
\end{equation}
The use of Eq.~(\ref{so}) in Eq.~(\ref{ges}) leads to
\begin{equation}
\left( \frac{d\phi}{d\varepsilon}\right)^2= \frac{U'(r)}{r \left[ 2U(r)-rU'(r) \right]}\, , \qquad
\left( \frac{dt}{d\varepsilon}\right)^2= \frac{2}{2U(r)-rU'(r)}\, .
\end{equation}

\begin{figure}[ht]
\centering
\subfigure[~Stability of the BH for the case $q=0$ and $\Lambda_\mathrm{eff}=0$  of Eq.~(\ref{con111})]{\label{fig:1a}
\includegraphics[scale=0.3]{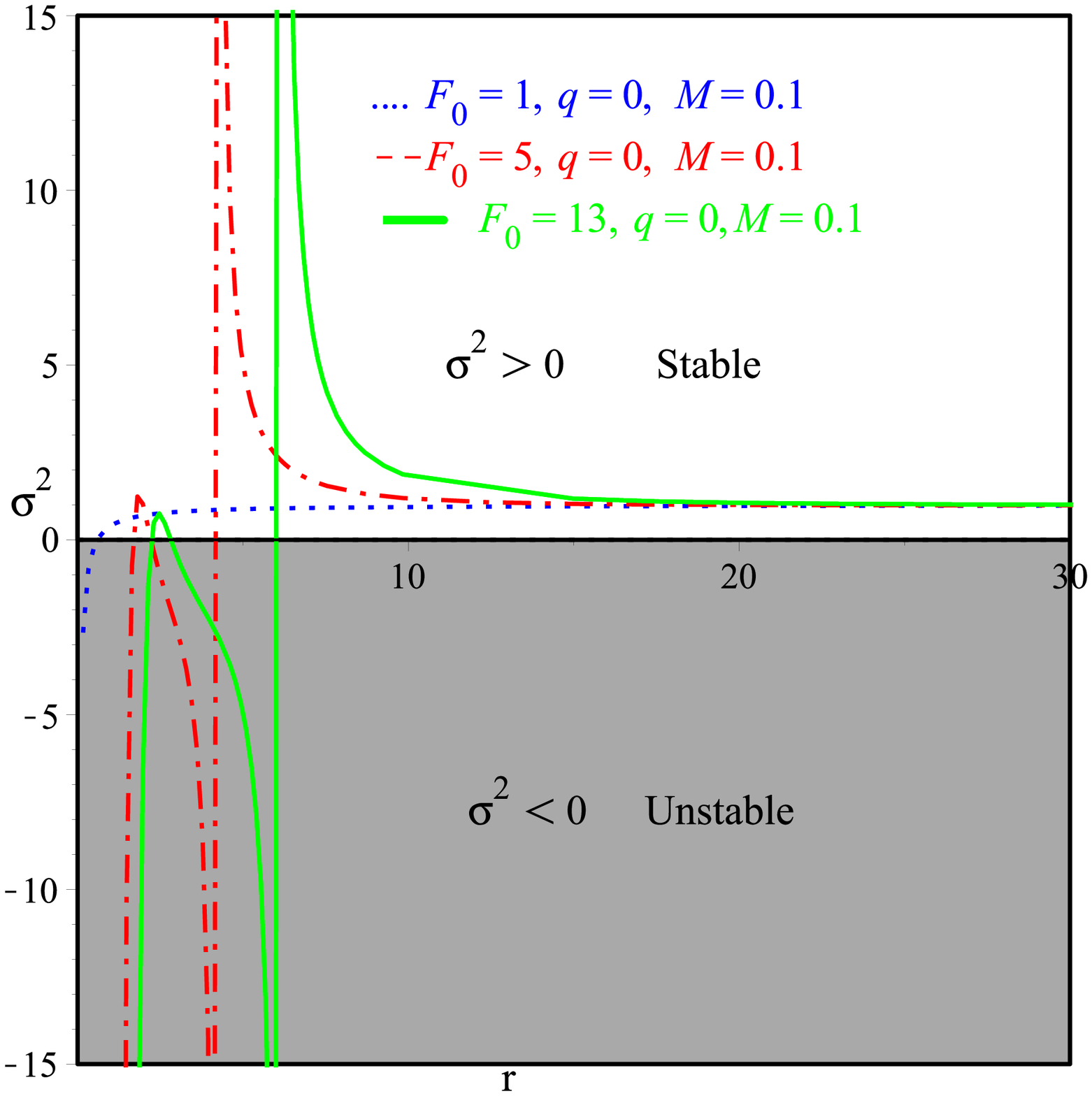}}\hspace{0.2cm}
\subfigure[~Stability of the BH for the case $q\neq 0$ and $\Lambda_\mathrm{eff}=0$ of Eq.~(\ref{con111})]{\label{fig:1b}
\includegraphics[scale=0.3]{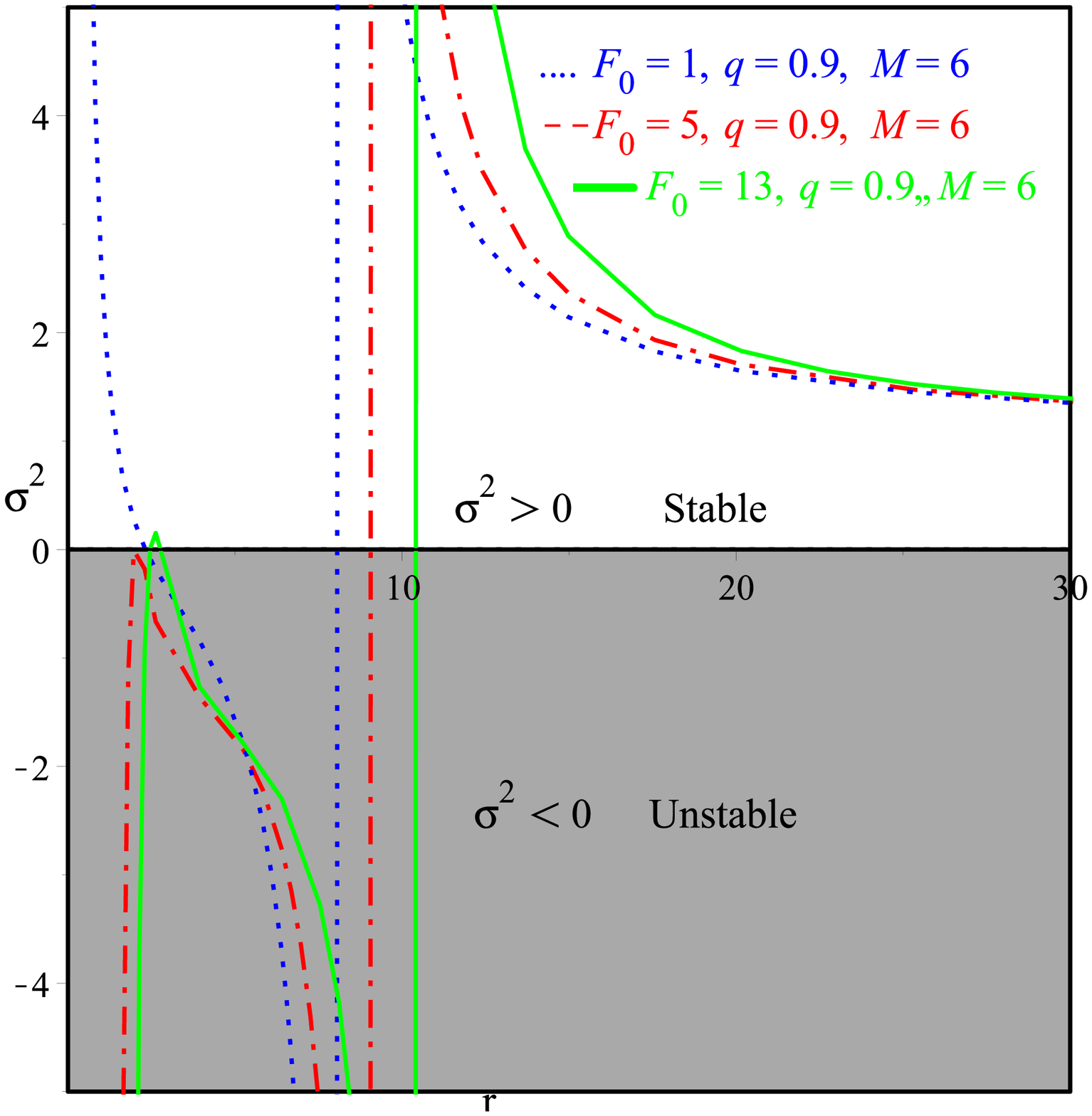}}
\caption{ {Plot of Eq.~(\ref{con111}) against the coordinate $r$ for the BHs (\ref{ass3n})
and (\ref{mpab}).}}
\label{Fig:1}
\end{figure}

We can rewrite Eq.~(\ref{ged11}) as
\begin{align}
\label{ged2222}
& \frac{d^2 \epsilon^1}{d\phi^2}
+U(r)U'(r) \frac{dt}{d\phi} \frac{d \epsilon^0}{d\phi}
 -2r V(r) \frac{d \epsilon^3}{d\phi} +\left[ \frac{1}{2} \left[U'(r)^2+U(r) U''(r)
\right]\left( \frac{dt}{d\phi}\right)^2-\left[U(r)+rU'(r) \right] \right]\zeta^1=0\, , \nonumber\\
& \frac{d^2 \epsilon^2}{d\phi^2}+\epsilon^2=0\, , \qquad
\frac{d^2 \epsilon^0}{d\phi^2} + \frac{U'(r)}{U(r)}
\frac{dt}{d\phi} \frac{d \epsilon^1}{d\phi}=0\, , \qquad \frac{d^2 \epsilon^3}{d\phi^2}
+ \frac{2}{r} \frac{d \epsilon^1}{d\phi}=0\, .
\end{align}
 From the second of Eq.~(\ref{ged2222}), we can show that we have a simple harmonic
motion, which is the stability condition of the plane $\theta=\pi/2$.
The remaining  equations of (\ref{ged2222}) assume the following solutions
\begin{equation}
\label{ged33}
\epsilon^0 = \zeta_1 \e^{i \sigma \phi}\, , \qquad
\epsilon^1= \zeta_2\e^{i \sigma \phi}\, , \qquad \mbox{and} \qquad
\epsilon^3 = \zeta_3 \e^{i \sigma \phi}\, ,
\end{equation}
where $\zeta_1, \zeta_2$, and $\zeta_3$ are constants and $\omega$ is an unknown {function}.
Using the values of $\epsilon^1$, and $\epsilon^3$ given by  Eq.~(\ref{ged33}) in the fourth equation of Eq.~(\ref{ged2222}), we get
\begin{equation}
\label{c1}
\zeta_2=-\frac{\zeta_3\sigma\,r}{2}\,.
\end{equation}
Then substituting the values of $\epsilon^0$, and $\epsilon^1$ given by Eq.~(\ref{ged33}) into the third equation of Eq.~(\ref{ged2222}) and using Eq.~(\ref{c1}),
we obtain
\begin{equation}
\label{c2}
\zeta_1=\frac{\zeta_3\sqrt{2U'\,r^3}}{2U}\,.
\end{equation}
Substituting (\ref{ged33}), after using Eqs.~(\ref{c1}) and (\ref{c2}) into the first equation of Eq.~(\ref{ged2222}),
we obtain the stability condition as
\begin{equation}
\label{con111}
\frac{3U V U'-\sigma^2 U U'-2rV{U'}^{2}+rUV U'' }{UU'}>0\Rightarrow \sigma^2<\frac{3U V U'-2rV{U'}^{2}+rUV U'' }{UU'}\, .
\end{equation}
We depict Eq.~(\ref{con111}) in Figure~\ref{Fig:1} using specific values of the model.
The case $q= 0$ is drawn in Figure~\ref{Fig:1}~\subref{fig:1a}, and the case
$q\neq 0$ is drawn in Figure~\ref{Fig:1}~\subref{fig:1b} for the  inequality of Eq.~(\ref{con111}). These  figures exhibit the unshaded and shaded zones where
the BHs are stable and not stable, respectively.

\section{Thermodynamics of the BH solutions (\ref{ass1}) and (\ref{ass1c})}\label{S5}


In this section, we will investigate the thermodynamical properties of the BH solutions (\ref{ass1}) and
(\ref{ass1c})\footnote{In this study, we will not address the BH solutions (\ref{ass1})and (\ref{ass1c})
because one cannot easily find their explicit roots. Therefore, we will use $\Lambda_\mathrm{eff}=0$
and study these solutions up to $O\left(\frac{1}{r^4}\right)$.}.
The temperature of a BH is defined as \cite{PhysRevD.86.024013,Sheykhi:2010zz,
Hendi:2010gq,PhysRevD.81.084040,Shirafuji:1997wy,Wang:2018xhw,Zakria:2018gsf}
\begin{equation}
\label{temp}
T_{(1,2)} =\frac{r_{(1,2)}-r_{(2,1)}}{4\pi r_{(1,2)}{}^2}\,,
\end{equation}
where $r_{(1,2)}$ represents the inner and outer horizons of the space-time.
The Hawking entropy of the horizons is defined as
\begin{equation}
\label{ent}
\psi_{(1,2)} =\frac{1}{4}A_{(1,2)}\,f_R\,,
\end{equation}
with $A_{(1,2)}$ being the area of the horizons.
The quasi-local energy is defined as \cite{PhysRevD.84.023515,PhysRevD.86.024013,
Sheykhi:2010zz,Hendi:2010gq,PhysRevD.81.084040,Zheng:2018fyn}
\begin{equation}
\label{en}
E_{(1,2)} =\frac{1}{4}\int \left[2f_R~r_{(1,2)}+r_{(1,2)}^2
\left\{F_0 \left( R \left( r_{(1,2)} \right) \right) - R \left(r_{(1,2)} \right)
f_R~r_{(1,2)}\right\}\right]dr_{(1,2)}\, .
\end{equation}
Finally, the Gibbs free energy is defined as \cite{Zheng:2018fyn,Kim:2012cma}
\begin{equation}
\label{enr}
G_{(1,2)}= E_{(1,2)}- T_{(1,2)}
{ \psi_{(1,2)}} \, .
\end{equation}


\subsection{Thermodynamics of the BH (\ref{mpab}) that has asymptotic flatness}

\begin{figure}
\centering
\subfigure[~The metric potentials of the BH (\ref{mpab})]{\label{fig:met}
\includegraphics[scale=0.22]{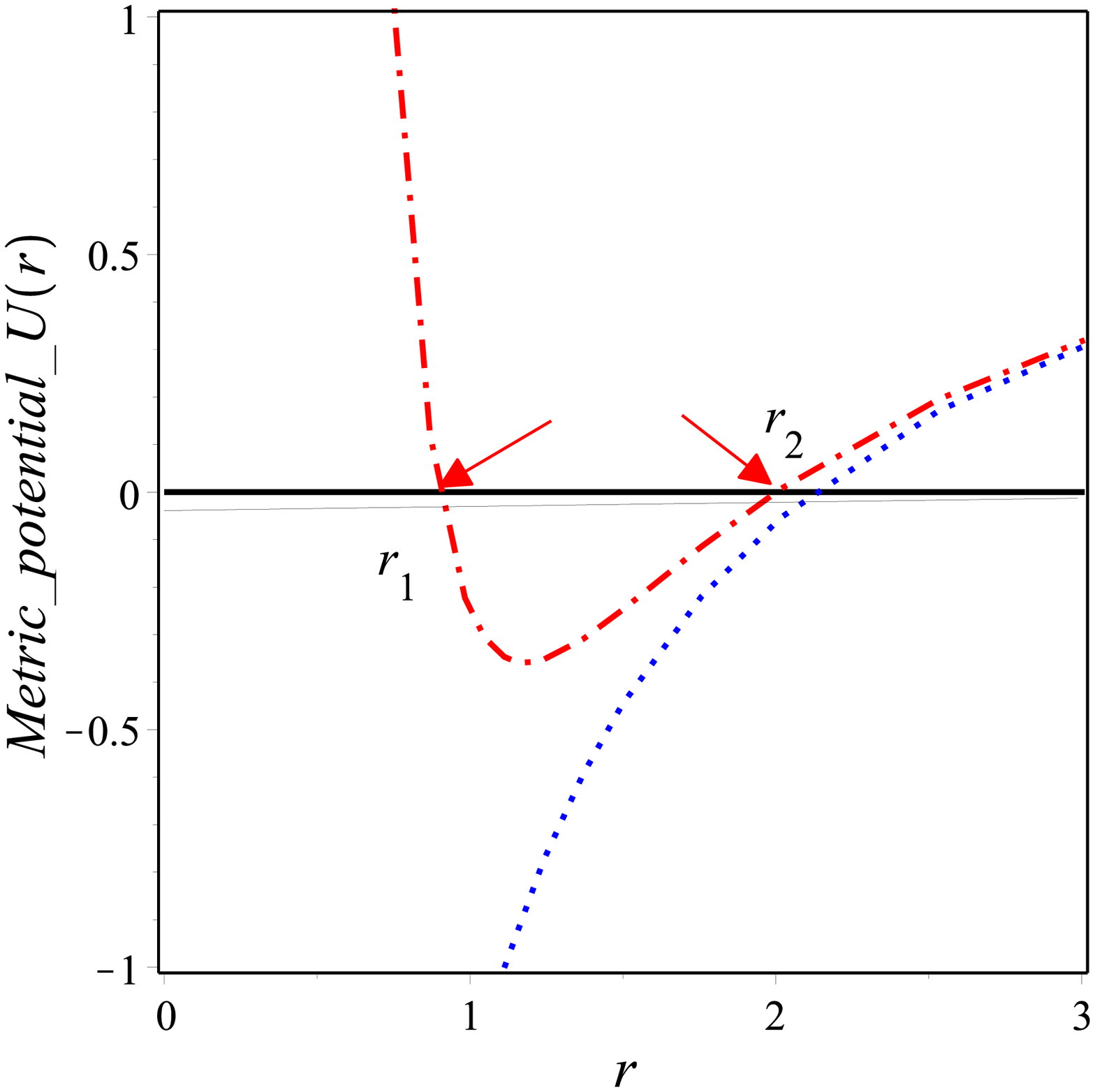}}
\subfigure[~The horizons of the BH (\ref{mpab}) of the metric potential $U$]{\label{fig:rd}
{\label{fig:metrd}\includegraphics[scale=0.22]{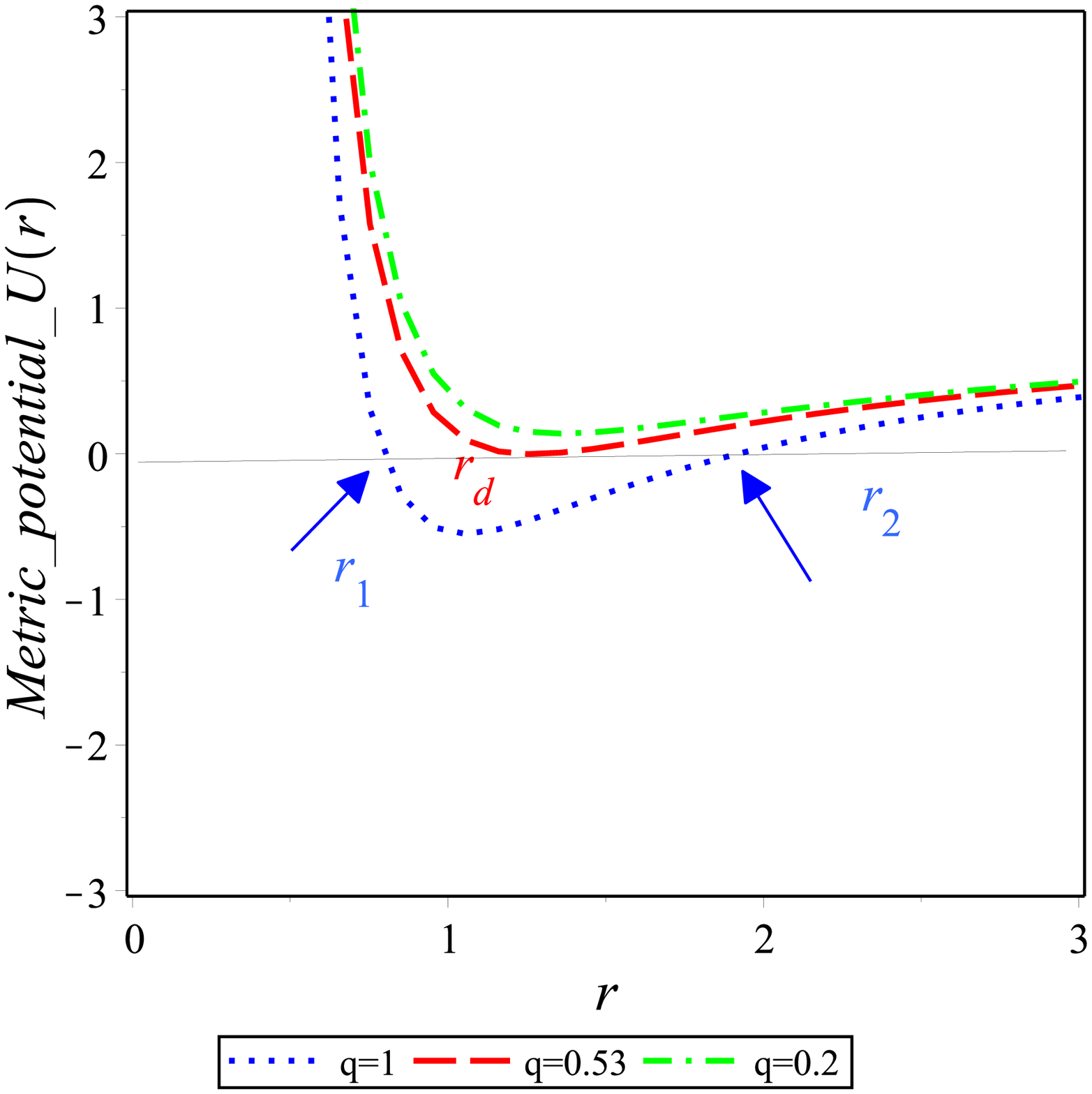}}}
\subfigure[~The mass of the BH (\ref{mpab}) when $F_0=1$]{\label{fig:mass}
\includegraphics[scale=0.22]{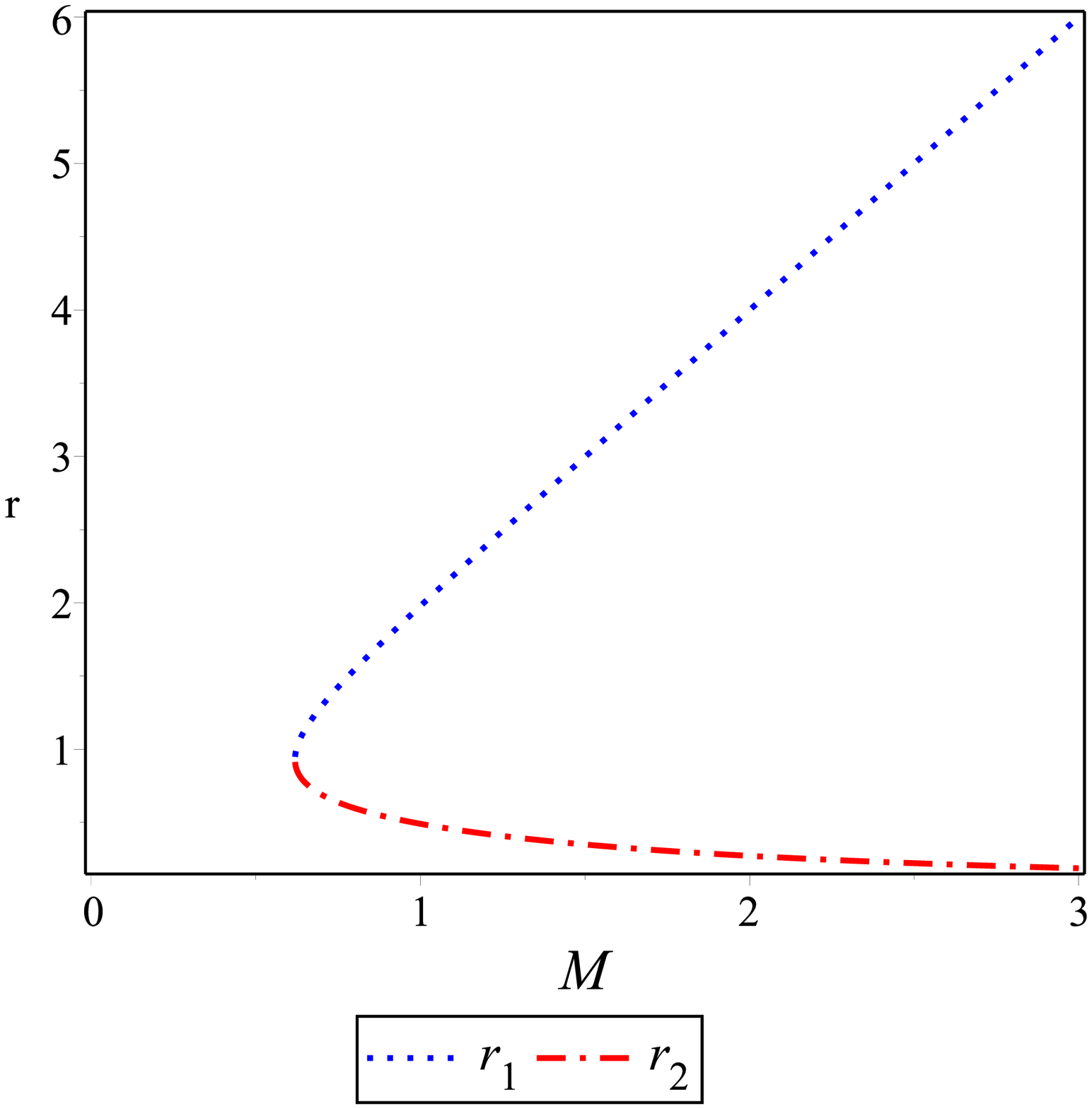}}
\subfigure[~The mass of the BH (\ref{mpab}) when $F_0=1.1$ ]{\label{fig:massfr}
\includegraphics[scale=0.22]{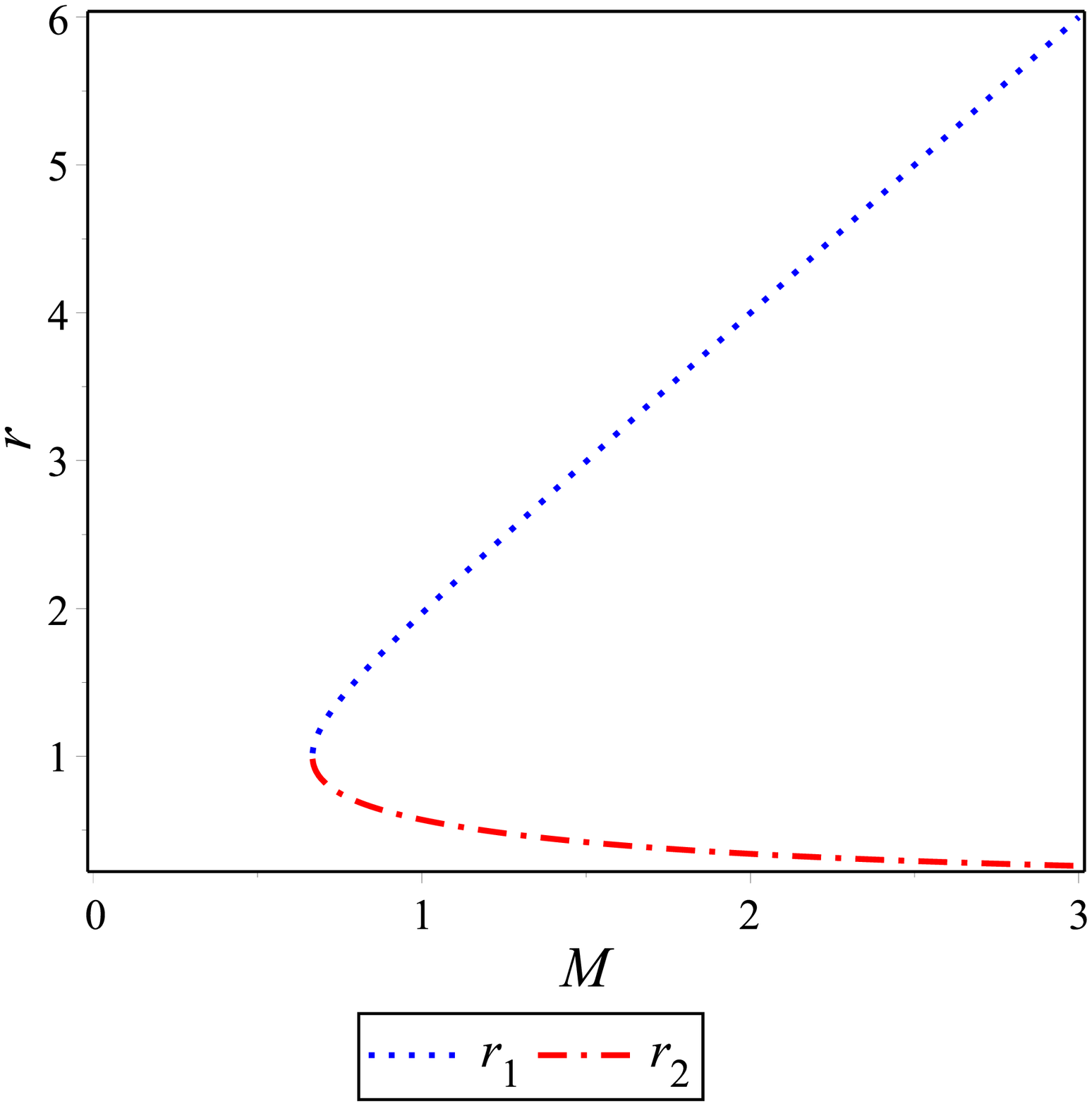}}
\subfigure[~Hawking temperature of the BH (\ref{mpab}) when $F_0=1$ ]{\label{fig:temp}
\includegraphics[scale=0.22]{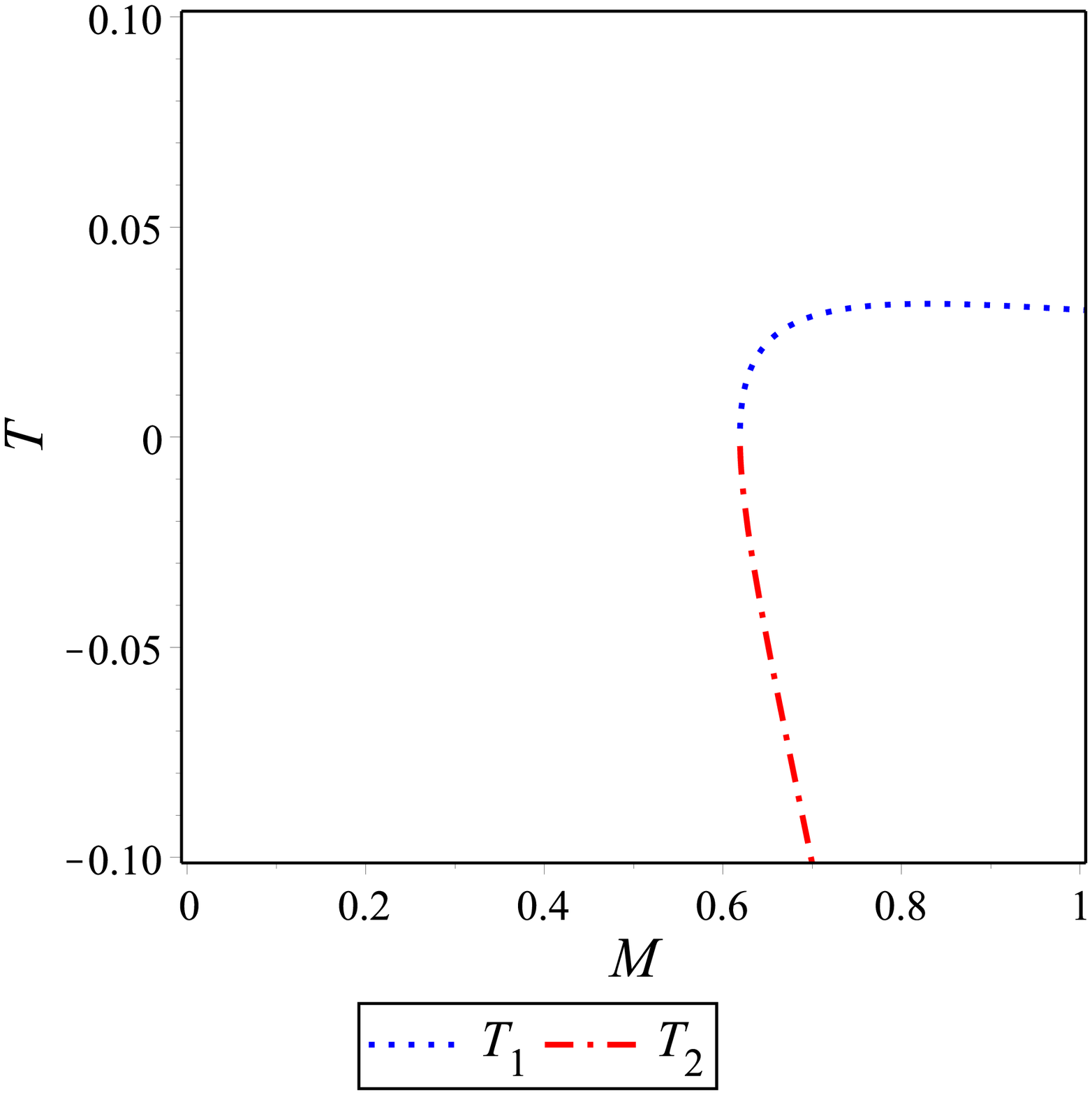}}
\subfigure[~Hawking temperature of the BH (\ref{mpab}) when $F_0=1.1$ ]{\label{fig:tempr}
\includegraphics[scale=0.22]{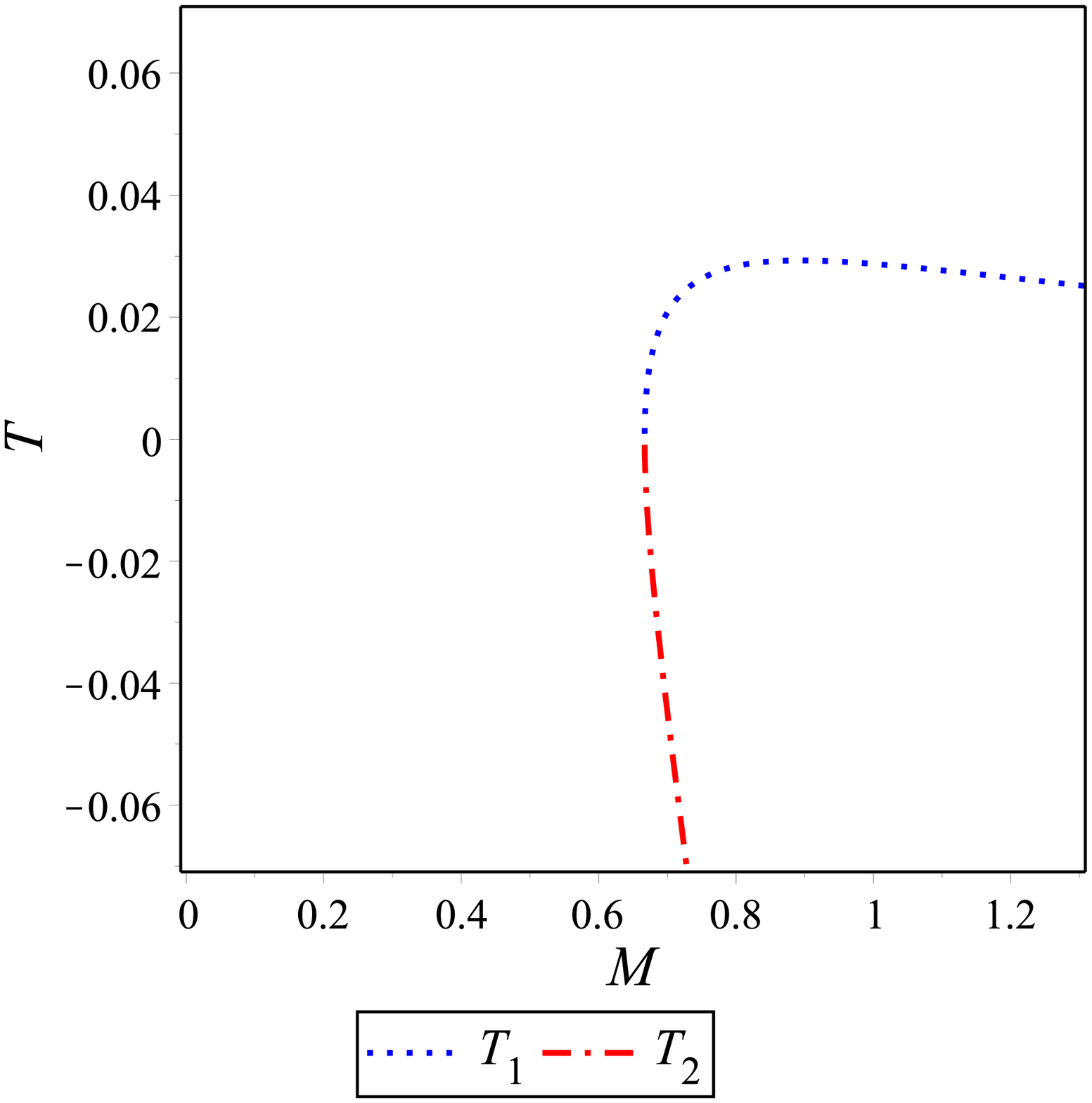}}
\subfigure[~The entropy of the BH (\ref{mpab}) when $F_0=1$ ]{\label{fig:ent}
\includegraphics[scale=0.22]{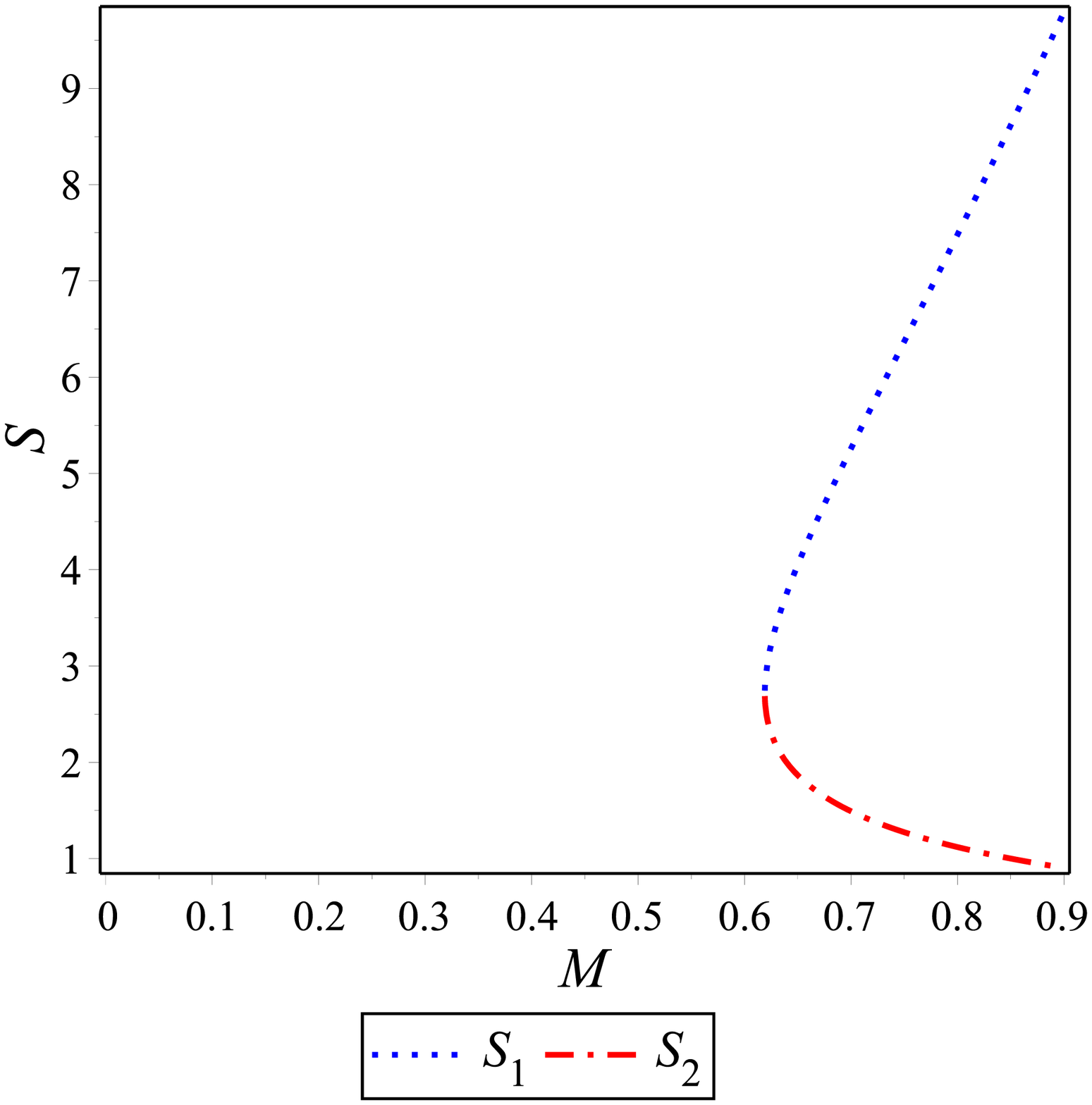}}
\subfigure[~The entropy of the BH (\ref{mpab}) when $F_0=1.1$ ]{\label{fig:entr}
\includegraphics[scale=0.22]{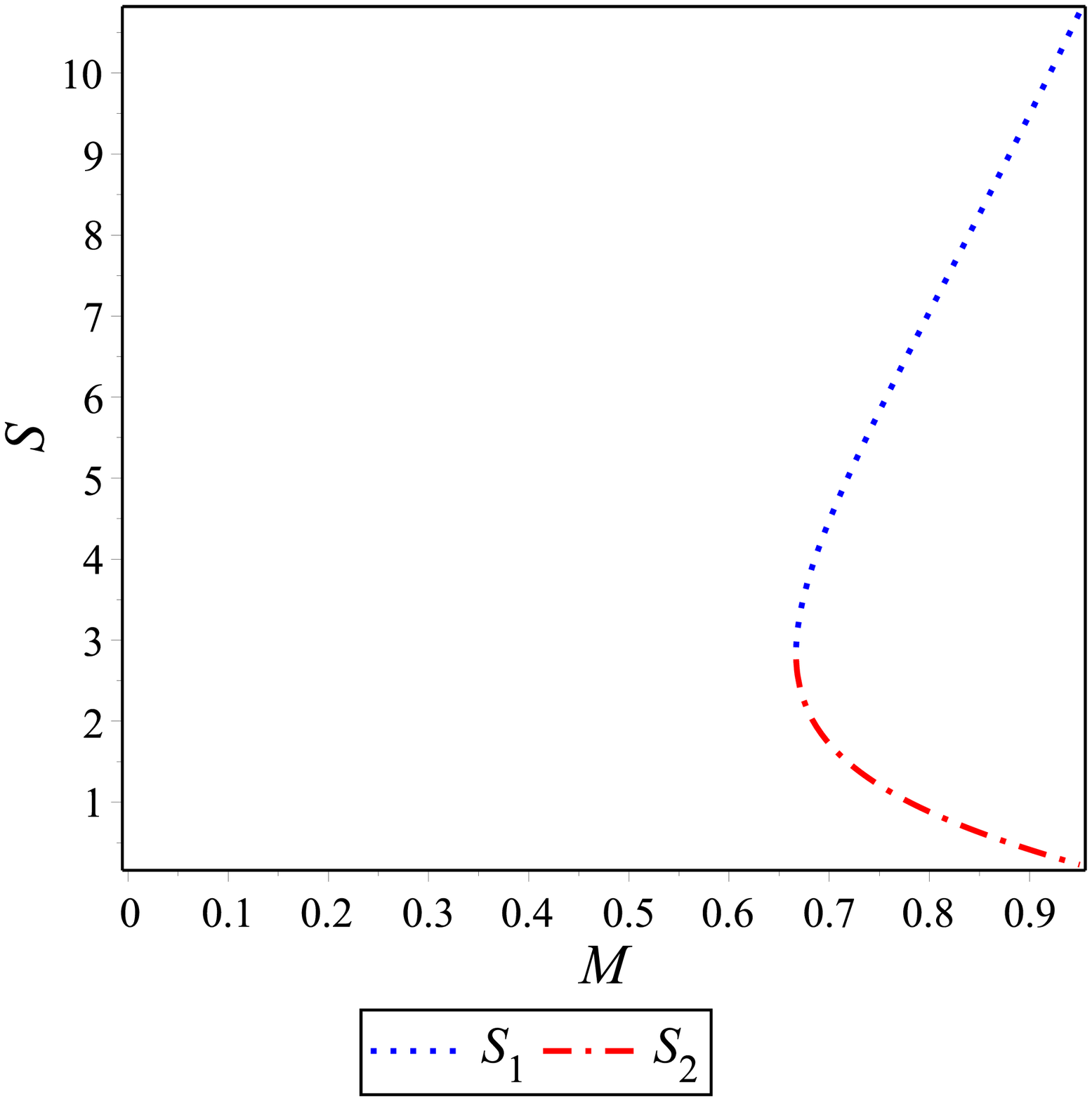}}
\subfigure[~The quasi-local energy of the BH (\ref{mpab}) when $F_0=1$ ]{\label{fig:enr}
\includegraphics[scale=0.22]{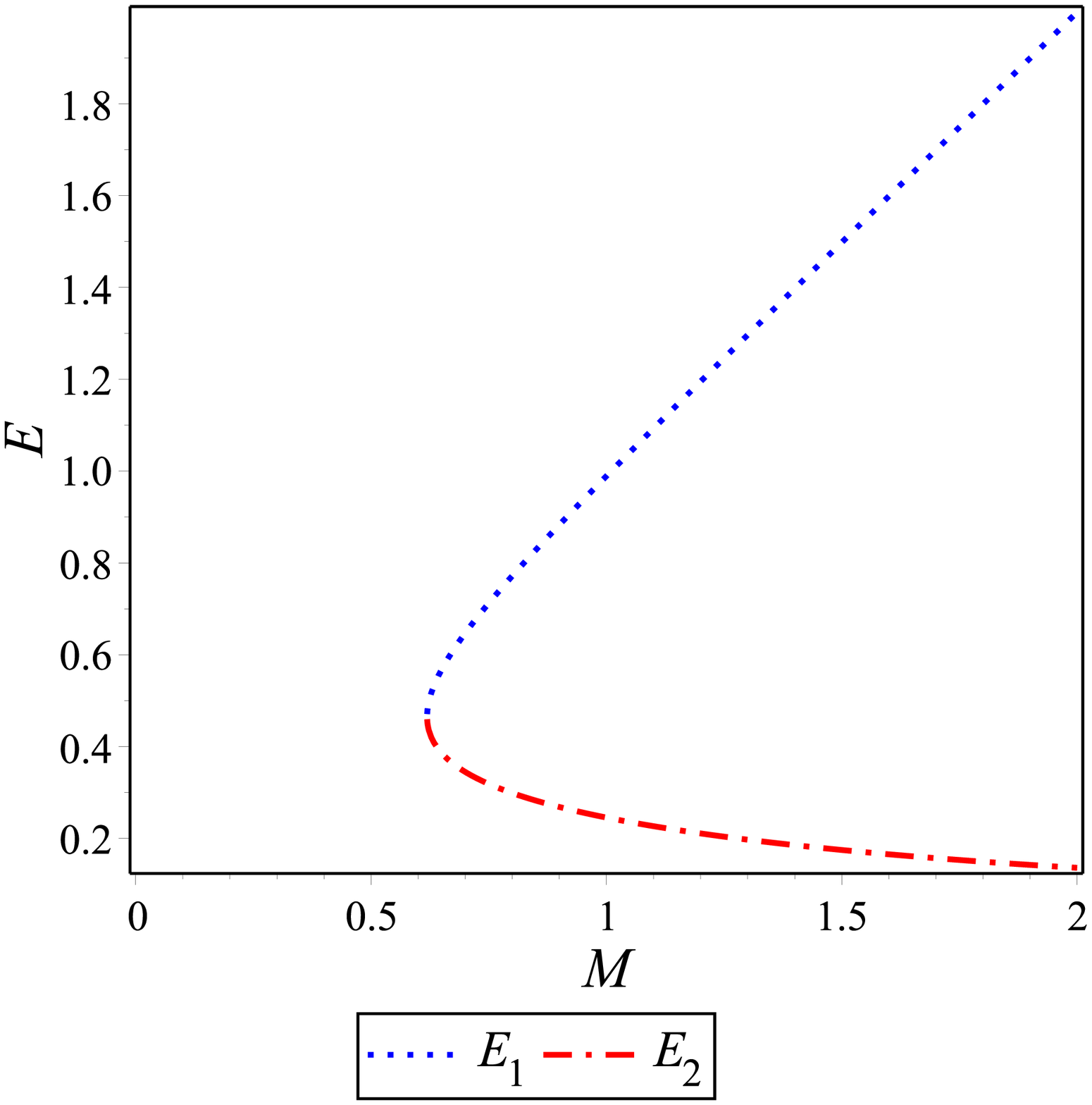}}
\subfigure[~The quasi-local energy of the BH (\ref{mpab}) when $F_0=1.1$ ]{\label{fig:enrr}
\includegraphics[scale=0.22]{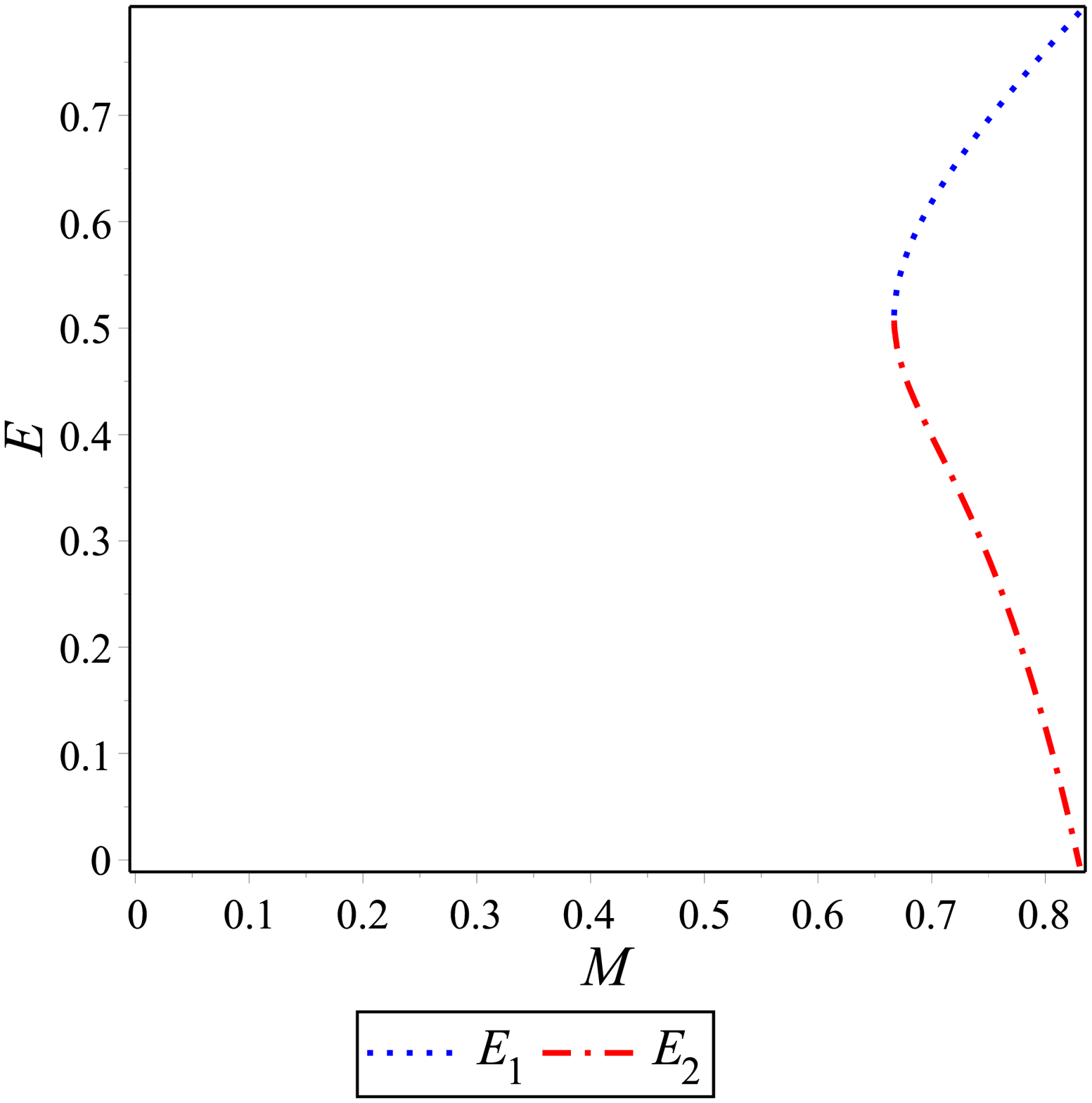}}
\subfigure[~The free energy of the BH (\ref{mpab}) when $F_0=1$ ]{\label{fig:gib}
\includegraphics[scale=0.22]{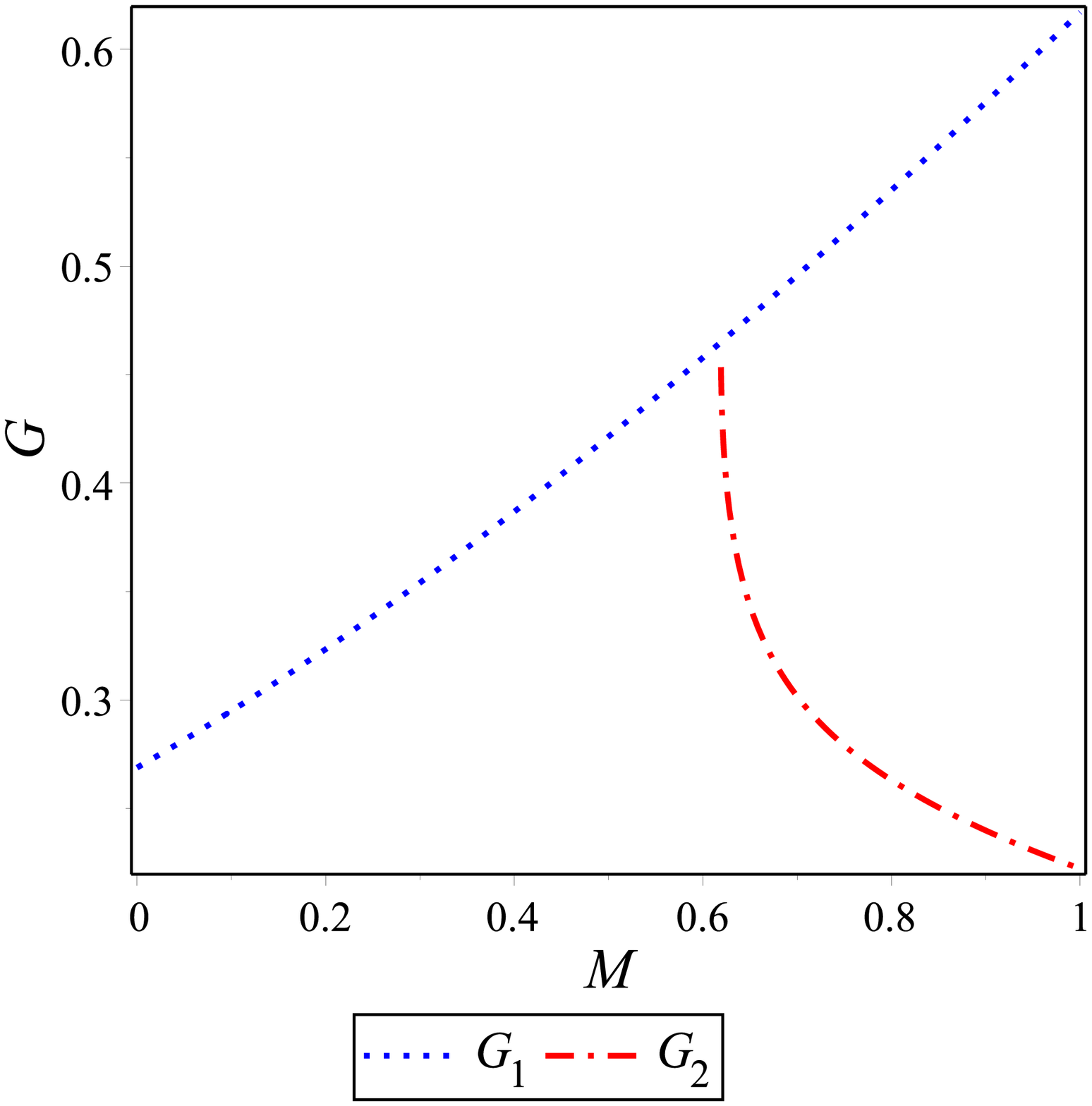}}
\subfigure[~The free energy of the BH (\ref{mpab}) when $F_0=1.1$ ]{\label{fig:gibfr}
\includegraphics[scale=0.22]{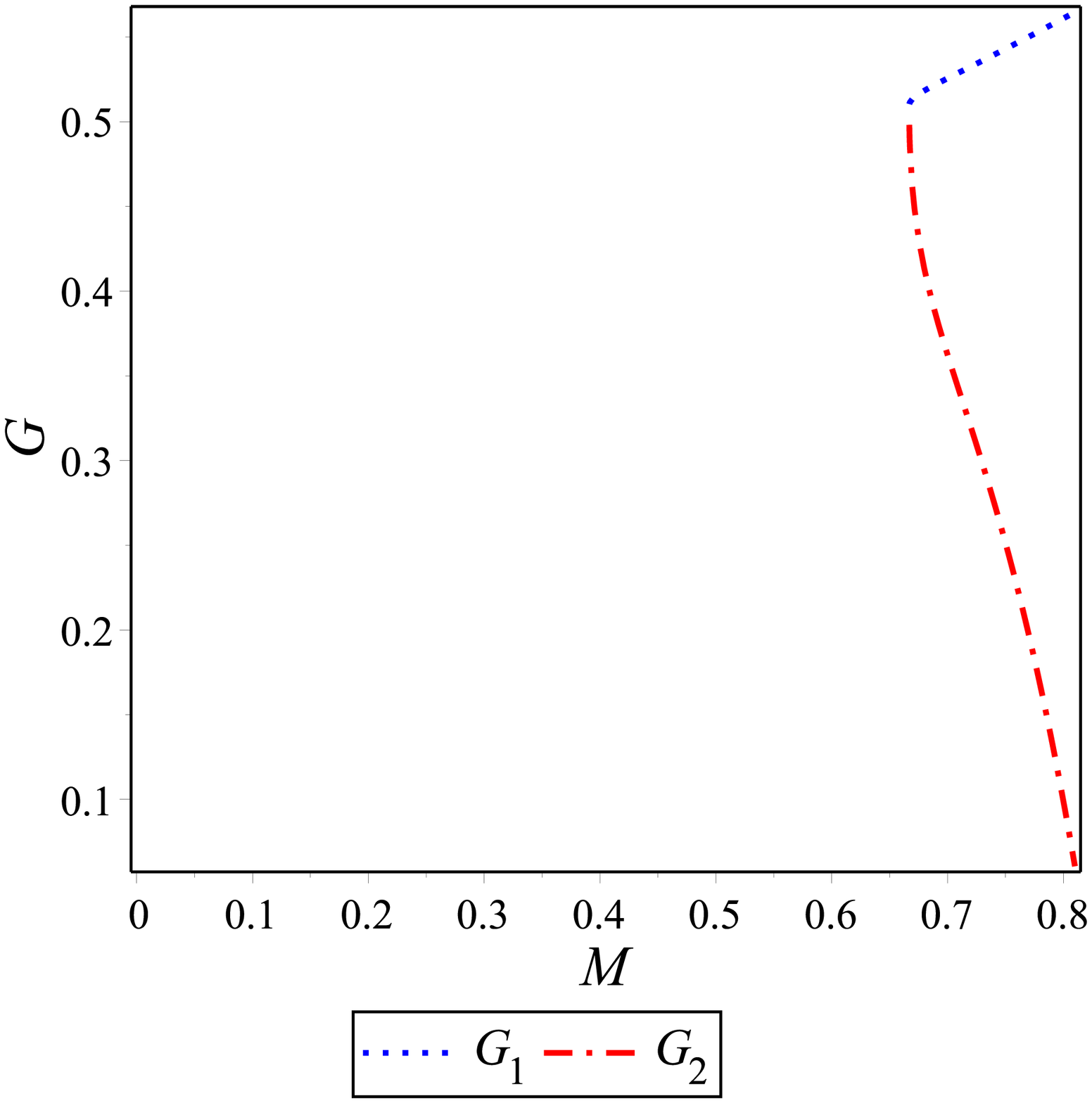}}
\caption[figtopcap]{Plots of thermodynamical quantities of the BH solution Eq.~(\ref{mpab}):
\subref{fig:met} shows the behavior of the metric potential $U$ for $F_0=1$ and $F\neq1$ when $m=1$ and $q=0.5$;
\subref{fig:metrd} Typical behavior of the horizons, degenerate and non- horizon of the metric potential $U(r)$ given by
Eq.~(\ref{mpab}) when $m=0.77$ and $F=2$;
\subref{fig:mass} Typical behavior of the horizons with mass given by
Eq.~(\ref{mpab}) when $q=0.05$ and $F_0=1$;
\subref{fig:massfr} Typical behavior of the horizons with mass given by
Eq.~(\ref{mpab}) when $q=0.05$ and $F=2$; \subref{fig:temp} and \subref{fig:tempr} The behavior of the Hawking temperature,
(\ref{temp}), showing $T_1$ has a positive decreasing value whereas $T_2$ has a negative increasing value when $F_0=1$ and $F_0=1.1$;
\subref{fig:ent} and \subref{fig:entr} The behavior of the Hawking entropy, (\ref{S1}), which indicates $\psi_{(1,2)}$
always has positive values, $ \psi_{1}$ and increases with $M$ while $\psi_{2}$ has a positive
decreasing value; \subref{fig:enr} and \subref{fig:enrr} The behavior of quasi-local energy, (\ref{E1}), which
indicate that $E_{(1,2)}$ has positive increasing values and also $E_1>E_2$; \subref{fig:gib} and \subref{fig:gibfr}
The behavior of Gibb's free energy which indicates the behavior of $G_1$ which is positive while $G_2$ starts
with a positive value that becomes negative as $M$ increases.}
\label{Fig:2}
\end{figure}

In this subsection, we study the thermodynamics of the BH (\ref{mpab}) which is
characterized by the mass $m$, $F$, and the constant $q$.
{ We only consider the asymptotically flat solution where $c_2=0$. }
When $F_0=1$, we obtain 
the Reissner-Nordstr\"om BH of GR. To derive the horizons of the BH,
(\ref{mpab}), we set $U(r)=0$ \cite{Zakria:2018gsf}, whose solutions give two real
positive roots that have the form
\begin{align}
\label{ged22}
r_1 =& \frac{1}{6} \left(3m+3\sqrt{\frac{q^4+12\left( F_0 - 1 \right)}{F_1}+m^2+\frac{2q^2}{3}+\frac{F_1}{9}} \right. \nonumber\\
& \left. +3\left\{2m^2
+\frac{4q^2}{3}-\frac{F_1}{9}-\frac{12\left( F_0 - 1 \right)}{F_1}-\frac{q^4}{F}+\frac{2mq^2(1+m)}{\sqrt{\frac{q^4+12\left( F_0 - 1 \right)}
{F_1}+m^2+\frac{2q^2}{3}+\frac{F_1}{9}}}\right\}^{1/2} \right)\,,\nonumber\\
r_2 =& \frac{1}{6}\left( 3m+3\sqrt{\frac{q^4+12\left( F_0 - 1 \right)}{F_1}+m^2+\frac{2q^2}{3}+\frac{F_1}{9}} \right. \nonumber\\
& \left. -3\left\{2m^2
+\frac{4q^2}{3}-\frac{F_1}{9}-\frac{12\left( F_0 - 1 \right)}{F_1}-\frac{q^4}{F}+\frac{2mq^2(1+m)}{\sqrt{\frac{q^4+12\left( F_0 - 1 \right)}
{F_1}+m^2+\frac{2q^2}{3}+\frac{F_1}{9}}}\right\}^{1/2}\right)\,,
\end{align}
where \\
\begin{align}
F_1=&3\left(6\sqrt{\left( 1 - F_0 \right) \left[18q^8+3m^2q^6+24\left( 1 - F_0 \right)q^4+108q^2m^2\left( 1 - F_0 \right) \right]
+48 \left(1-3F_0 -3F_0^2 \right)} \right. \nn
& \left. -q^6+36q^2\left( F_0 - 1 \right)+54m^2\left( F_0 - 1 \right)\right)^{1/3}\,.\nonumber
\end{align}
The metric potentials of the BH (\ref{mpab}) are plotted in Figure~\ref{Fig:2}
\subref{fig:met}. Figure~\ref{Fig:2} \subref{fig:met} shows that we have two positive horizons of
the metric potential $U(r)$ when $F\neq1$. The degenerate horizon of $U(r)$ occurred at
$(F,m,q,r)\equiv(2,0.77,0.53,1.22)$
{ where the radii of the two horizons coincides with each other, $r_1=r_2$. }
The behavior of the degenerate is indicated in Figure~\ref{Fig:2} \subref{fig:metrd} which shows that $r_1<r_d$.
As we observe from Figure~\ref{Fig:2} \subref{fig:metrd}, as $q$ decreases, we enter a parameter region
where no horizon exists, and thus, the central singularity { becomes} 
a naked singularity.
The behaviors of the two horizons w.r.t. mass are drawn in Figure~\ref{Fig:2} \subref{fig:mass}
and \subref{fig:massfr} when $F_0=1$ and $F_0\neq1.1$.
{ The interesting point is that the mass is not equal to the absolute value of the charge $m\neq \left| q \right|$
in the extremal limit where the radii of the two horizons coincide with each other, $r_1=r_2$.
In the case of the Reissner-Nordstr\"om BH of GR, the mass equals the absolute value of the charge $m=\left| q \right|$.
As clear from Eq.~(\ref{temp}), the Hawking temperature vanishes in the extremal limit.
}

From Eq.~(\ref{temp}) the Hawking temperature of the BH (\ref{mpab}) is calculated and drawn
in Figs.~\ref{Fig:2} \subref{fig:temp} and \subref{fig:tempr} for $F_0=1$ and $F_0=1.1$, respectively.
These figures show that $T_{1}>T_{2}$. Figs.~\ref{Fig:2} \subref{fig:temp} and \subref{fig:tempr}
also indicate that $T_1$ has an increasing positive value whereas $T_2$ has a decreasing negative one.

 From Eq.~(\ref{ent}), the entropy of BH (\ref{mpab}) takes the form
\begin{eqnarray}
\label{S1}
\psi_{(1,2)}= \pi\, r_{(1,2)}{}^2\left(1-\frac{F_0-1}{r_{(1,2)}{}^4}\right)\,.
\end{eqnarray}
The plot of the entropy (\ref{S1}) when $F_0=1$ and $F_0=1.1$ is drawn in Figs.~\ref{Fig:2} \subref{fig:ent}
and \ref{Fig:2} \subref{fig:entr} which indicate an increasing value for $\psi_1$ and decreasing value for $\psi_2$.
 From Eq.~(\ref{en}), we evaluate the local energy of BH (\ref{mpab}) and obtain
\begin{eqnarray}
\label{E1}
E_{(1,2)}=\frac{21{r_{(1,2)}}^9+\left( F_0 - 1 \right) \left[ 7{r_{(1,2)}}^5-126q^2 {r_{(1,2)}}^3+54r_{(1,2)}
\left( F_0 - 1 \right)-126m\left( F_0 - 1 \right)\right]}{42{r_{(1,2)}}^8}\,.
\end{eqnarray}
Equation (\ref{E1}) shows that when $F_0=1$, we obtain $E_{(1,2)}=\frac{r_{(1,2)}}{2}$
which is the energy of a spherically symmetric space-time.
The plot of Eq.~(\ref{E1}) when $F_0=1$ and $F_0=1.1$ is drawn in Figs.~\ref{Fig:2} \subref{fig:enr}
and \ref{Fig:2} \subref{fig:enrr}, which also indicates positive increasing values for $E_{(1,2)}$.
Figs.~\ref{Fig:2} \subref{fig:enr} and \subref{fig:enrr} also show
$E_1>E_2$. Finally, by using of Eqs.~(\ref{temp}), (\ref{S1}), and (\ref{E1}) in (\ref{enr}), we
evaluate Gibbs' free energies. The plot of these energies is drawn in Figure~\ref{Fig:2}
\subref{fig:gib} and \ref{Fig:2} \subref{fig:gibfr} when $F_0=1$ and $F_0=1.1$.
These figures indicate a positive increasing value for $G_1$ and also $G_1>G_2$.


\subsection{First law of thermodynamics of the BH solutions (\ref{ass3n}) and (\ref{mpab})}\label{fir}


An important step for any BH solution is to check its validity of the first law of thermodynamics.
Therefore, for the charged BH the Smarr formula and the differential form for the first law of
thermodynamics, in the frame of the $f(R)$ gravity, can be expressed as \cite{Zheng_2018,Okcu:2017qgo}
\begin{equation}
\label{1st}
M=2 \left( T\psi-PV \right)+\eta Q\,, \qquad \qquad dE=Td\psi+\eta dQ+PdV\, ,
\end{equation}
where $\psi$ is the Hawking entropy, $T$ is the Hawking
temperature, $\eta$ is the electric potential $P$ is the radial component of the stress-energy tensor that is used as a
thermodynamic pressure, i.e., $P=T_r{}^r\mid_{\pm}$ , $V$ is the geometric volume and $M=m$, $Q=q$. The
pressure, in the context of the $f(R)$ gravity, is determined as \cite{Zheng_2018}
\begin{equation}
P=-\frac{1}{8\pi}\left\{\frac{F_1}{{r_{(1,2)}}^2}
+\frac{1}{2} \left( f(R)-RF_1 \right)\right\}
+\frac{1}{4}\left(\frac{2F_1}{r_{(1,2)}}+F'_1\right)T\,.
\label{2nd}
\end{equation}
Using Eq.~(\ref{metaf}), we get
\begin{equation}
M=\frac{1}{2}\left(r_{(1,2)}+\frac{q^2}{r_{(1,2)}}-\frac{\left( F_0 - 1 \right)}{2{r_{(1,2)}}^3}\right)\,.
\end{equation}
By calculating the necessary components of Eq.~(\ref{1st}), we obtain
\begin{align}
\label{1stf}
P_{(1,2)}\approx& -\frac{q^2{r_{(1,2)}}^2-3\left( F_0 - 1 \right)}{6{r_{(1,2)}}^3}\,,\qquad \qquad
T_{(1,2)}\approx \frac{{r_{(1,2)}}^4-q^2{r_{(1,2)}}^2+3\left( F_0 - 1 \right)}{4\pi {r_{(1,2)}}^5}\,, \nonumber \\
\psi_{(1,2)}\approx &\frac{\pi \left[{r_{(1,2)}}^4-\left( F_0 - 1 \right) \right]}{{r_{(1,2)}}^2} \,, \qquad \qquad
\eta_{(1,2)}\approx \frac{q}{2r_{(1,2)}}-\frac{\left( F_0 - 1 \right)q}{{r_{(1,2)}}^5}\,, \qquad \qquad
V=\frac{4}{3}\pi {r_{(1,2)}}^3.
\end{align}
Using Eq.~(\ref{1stf}) in Eq.~(\ref{1st}), we can prove that the first law of the flat space-times
(\ref{ass3n}) and (\ref{mpab}) is verified.

\section{Discussion and conclusions}\label{S77}

This study focuses on deriving spherically symmetric BH solutions with/without charge
in the context of the $f(R)$ modified gravity theory.
The $f(R)$ gravity is known as a theory whose field equation includes fourth order derivative
thereby making the derivation of the analytic solution difficult.
Therefore, we used the trace equation of the $f(R)$ equation and solved it with respect to $f(R)$.
Using this solution, we rewrote the charged field equations of the $f(R)$ gravity and applied them
to a spherically symmetric space-time that has two unknown functions of radial coordinate.
The resulting non-linear differential equations were solved under two cases:
The case of vanishing the electric charge and the case of non-vanishing electric charge.
In these two cases, we assumed the derivative of the $F(R)$,
$F_1(r)=\frac{df(R(r))}{dR(r)}=1-\frac{F_0-\left(n-3\right)}{r^n}$ where $n>0$.
{ By using }
the previous mentioned assumption that when $F_0=n-3$, we obtained $F_1(r)=1$, which is the case of GR.
In the frame of the above assumption of $F_1(r)$, we solved the field equations of the $f(R)$ gravity with/without the electric charge analytically
and derived the exact form of the metric potentials and the electric charge.

To understand the physics of these BH solutions, we gave the asymptotic form of the metric potentials, when $n=4$, with/without
charge. From such asymptote, we have shown explicitly that our BH solutions are different from GR BH solutions and coincide with them
when $F_0=1$.
Then we wrote the line-element of such BHs and have shown that they asymptotically behave as (A)dS space-time.
We calculated the asymptotic form of $f(R)$
{ when $r$ is large and have shown that $f(R)$ could include the terms with the positive power of $R$ higher than the Einstein- Hilbert term
and terms with negative power. }
We also calculated the {curvature} invariants of these BHs with/without charge and
have showm that the singularity of the Ricci squared tensor in the charged case is milder than that of the GR BHs.
Moreover, we studied the geodesic deviations of these BHs and derived the condition of stability analytically
and graphically, as shown in Figure~\ref{Fig:1}.

To further examine the physics of these BHs we 
{ considered} the thermodynamical quantities such as Hawking temperature, entropy, quasi-local energy, and Gibb's
free energy, and investigate their behavior analytic and graphically.
 In the case of the asymptotically flat solution, there exists the extremal limit,
where the radii of the two BH horizons, that is, inner horizon and outer horizon, coincide with each other and
the Hawking temperature vanishes.
In the case of the Reissner-Nordstr\"om BH of GR, the mass equals to the absolute value of the charge in the extremal limit.
It could be interesting that in our model, the mass is not equal to the absolute value of the charge in the limit,  where the radii of the two horizons coincide with each other.
Because the Hawking temperature vanishes, it is often considered that the extremal limit might be the remnant after the BH evaporation by the Hawking radiation.
Then it may give any clue for solving the information loss problem in the black hole because the disagreement of the mass and the charge may include some information is lost in GR. 
Moreover, we tested the first law of thermodynamics and have shown in detail that such BHs satisfy this law.
If we apply the results of odd perturbation presented in \cite{Elizalde:2020icc}, we can prove that the BHs presented in
this study are stable.

In summary, we derived new BHs with/without charge in the frame of $f( R )$ and showed that their
Ricci scalars are not constant. These BHs are original ones, and their originality comes from the constant that
involves, $F_0$. This constant comes from the assumption of the first derivative of $f(R)$, i.e.,
$F_1(r)=\frac{df(R(r))}{dR(r)}=1-\frac{F_0-\left(n-3\right)}{r^n}$ .

\begin{acknowledgments}
This work is supported by the JSPS Grant-in-Aid for Scientific Research (C)
No. 18K03615 (S.N.).
\end{acknowledgments}

%

\end{document}